\documentclass[amssymb, nobibnotes, aps,pra,preprint]{revtex4-1}
\usepackage{graphicx,epsfig,epstopdf}
\usepackage{amsmath} 
\usepackage{bm}
\usepackage{color}
\usepackage{subfig}

\begin{document}
\title{Detecting Electronic Coherence by Multidimensional Broadband Stimulated X-Ray Raman Signals}
\author{Konstantin E. Dorfman}
\email{kdorfman@uci.edu}
\author{Kochise Bennett}
\author{Shaul Mukamel}
\affiliation{Department of Chemistry, University of California, Irvine,
California 92697-2025, USA}
\date{\today}%

\begin{abstract}
Nonstationary molecular states which contain electronic coherences can be impulsively created and manipulated by using recently-developed ultrashort optical and X-ray pulses via photoexcitation,  photoionization and Auger processes. We propose several stimulated-Raman detection schemes that can monitor the phase-sensitive electronic and nuclear dynamics. Three detection protocols of an X-ray broadband probe are compared - frequency dispersed transmission, integrated photon number change, and total pulse energy change. In addition each can be either linear or quadratic in the X-ray probe intensity. These various  signals offer different gating windows into the molecular response which is described  by correlation functions of electronic polarizabilities. Off-resonant and resonant signals are compared.
\end{abstract}

\maketitle

\vspace{0.5cm}
\section{Introduction}
Many spectroscopic techniques involve the creation and manipulation of coherences followed by  a stimulated Raman detection  of a probe pulse\cite{Pes07, li2008two,Xin02,Mul02,Kee04,Vol05,Vac07,Wal09,Wan10,Roy10,Dor11,farrell2012ultrafast}. The field of multidimensional spectroscopy was launched by looking at the delays between impulsive stimulated Raman events \cite{tan93}. In the simplest conventional one-dimensional (1D) time-domain stimulated Raman technique, the molecule is first  prepared in a vibrational superposition state by an off-resonant Raman pulse and, after a variable delay period  $T$, the transmission change of a second probe pulse is detected. The transmission oscillates with $T$ between gain and loss at the vibrational period, and a Fourier transform then reveals the vibrational frequencies \cite{eld11,dor132,silberberg2014}. Optical Raman techniques have been applied to study electron transfer and nonadiabatic dynamics at conical intersections. Using recently developed FEL and HHG sources \cite{manzoni2014coherent,shwartz2014x,popmintchev2012bright}, Raman techniques can be further extended to the X-ray regime \cite{tanaka2002coherent,biggs13,Hudock:JPCA:2007,miyabe2015transient,mcfarland2014ultrafast} whereby the system is initially prepared in the superposition of valence electronic states and an X-ray Raman probe then reveals information about electronic, rather than vibrational, coherence.
\par
We had recently investigated multidimensional Raman techniques in the visible/IR regime \cite{dor132}. Signals quadratic in the probe pulse intensity which use broadband or shaped pulses to probe excited state dynamics following the preparation by resonant UV actinic pulse were studied. Various simulation protocols which differ by the level of complexity and computation cost were compared. Starting with the most computationally expensive expensive exact numerical propagation, we developed more approximate sum-over-states expansion and the Stochastic Liouville Equation (SLE) to describe complex dynamics. Excited state preparation has been limited to excited electronic state populations which do not carry phase information. In our recent work \cite{aga15} we had extended the SLE approach to both linear and quadratic off-resonant Raman signals with shaped pulses.  The Raman resonance line shapes were calculated for systems with nonequilibrium initial state prepared by a resonant actinic pulse.  We then investigated the relative phase dependence of time evolving Stokes and anti-Stokes components of the spectra in quadratic signal with shaped pulse \cite{batignani2015energy}. The results show the energy redistribution between different spectral components in Raman signals.
\par
This paper presents several electronic Raman techniques performed with X-ray pulses that carry higher levels of information \cite{biggs,biggs13,zha14}.  The same techniques apply to conventional vibrational Raman as well. Using an intuitive diagrammatic approach, we derive expressions for three basic detection protocols of the probe: $(i)$ the change in number of photons $S^{(N)}$, $(ii)$ the change in the transmitted probe energy  $S^{(E)}$ and $(iii)$ the frequency-dispersed probe transmission $S^{(fd)}$. In addition, signals may be linear or quadratic in the probe intensity and may utilize broadband or hybrid shaped (a combination of broad and narrow) probes as well as resonant or off-resonant with material transition. The energy exchange between field and matter is discussed. In the case of an off-resonant probe, the number of photons is conserved and $S^{(N)}=0$. For a hybrid pulse composed of a narrowband and a broadband component, we see oscillations between Stokes/anti-Stokes components \cite{batignani2015energy} with the delay $T$, whereas for a single broadband pulse the entire pulse envelope oscillates, smearing out the spectral features in the hybrid signal \cite{ide13}.
\par
We present a systematic organization scheme for stimulated Raman multidimensional spectroscopies applicable when the preparation of some nonstationary state, the nature or dynamics of which is the object of study, is temporally well-separated from the detection process.  Many conventional spectroscopic techniques (e.g. pump-probe type) fulfill these conditions \cite{bostedt2013ultra,katayama2013ultrafast,lemke2013femtosecond}. The scheme takes the inital state as a given and classifies signals by their dependence on detection parameters.  These parameters control which features of the nonstationary state are observable and how they manifest.   In particular, we examine different choices for the field (whether the field spectrum is resonant or off-resonant with the material transitions), the intensity scaling with the detecting field (linear or quadratic), the spectral shape of the detecting field (broadband or hybrid broad-narrowband).  These are only a few of the possible combinations of parameters that could describe spectroscopic detection of a nonstationary state.  Similar ideas were pursued for off-resonant X-ray scattering (diffraction) \cite{bennett2014time, biggs2014multidimensional} and for spontaneous emission following impulsive Raman X-ray excitation \cite{dorfman2013nonlinear}.
\par
While off-resonant Raman signals are simpler to analyze, resonant pulses are more selective to a given atomic core transition and provide additional specific information about molecules. We examine different simulation protocols and derive expressions that may be used for direct numerical propagation of the wavefunction in Hilbert space, which includes all degrees of freedom explicitly, or the density matrix in Liouville space \cite{fin14},  which gives a simple picture of bath effects and allows for a reduced description via the Stochastic Liouville Equations (SLE) \cite{dor132, and14,aga15}.
\par
In the off-resonant regime the field-matter interaction Hamiltonian is the product of field intensity and a molecular polarizibility, while in the resonant regime it is the dot product of field amplitude and molecular transition dipole. The off-resonant signals are explored first and expressions for each of the three detection protocols for signals linear and quadratic in the field intensity are derived.  We then present plots of these signals for a simple model system and discuss the properties of each signal and what information it reveals.  This analysis is then repeated for the resonant signals using the same model.  We conclude with a comparison of off-resonant versus resonant signals and a general discussion of the utility of this signal classification scheme.

\section{Detection Protocols for Off-Resonant Raman signals}\label{sec:OffRes}

\begin{figure}[h]
\centering
\includegraphics[width=.2\textwidth]{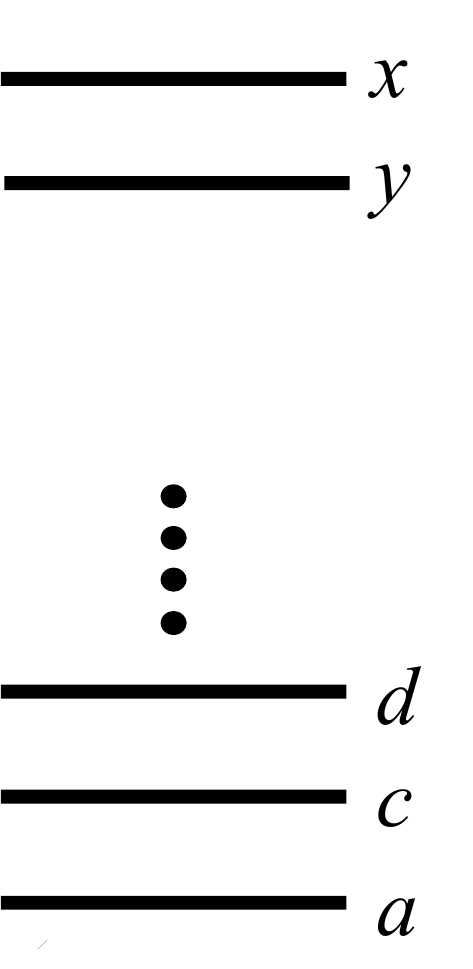}
\caption{Level scheme for the broadband X-ray Raman experiment. $a$, $c$, $d$ are low-frequency, valence electronic excitations and $x$, $y$ are high-frequency core excitations.}
\label{fig:scheme}
\end{figure}
\begin{figure}
\centering
\includegraphics[width=.85\textwidth]{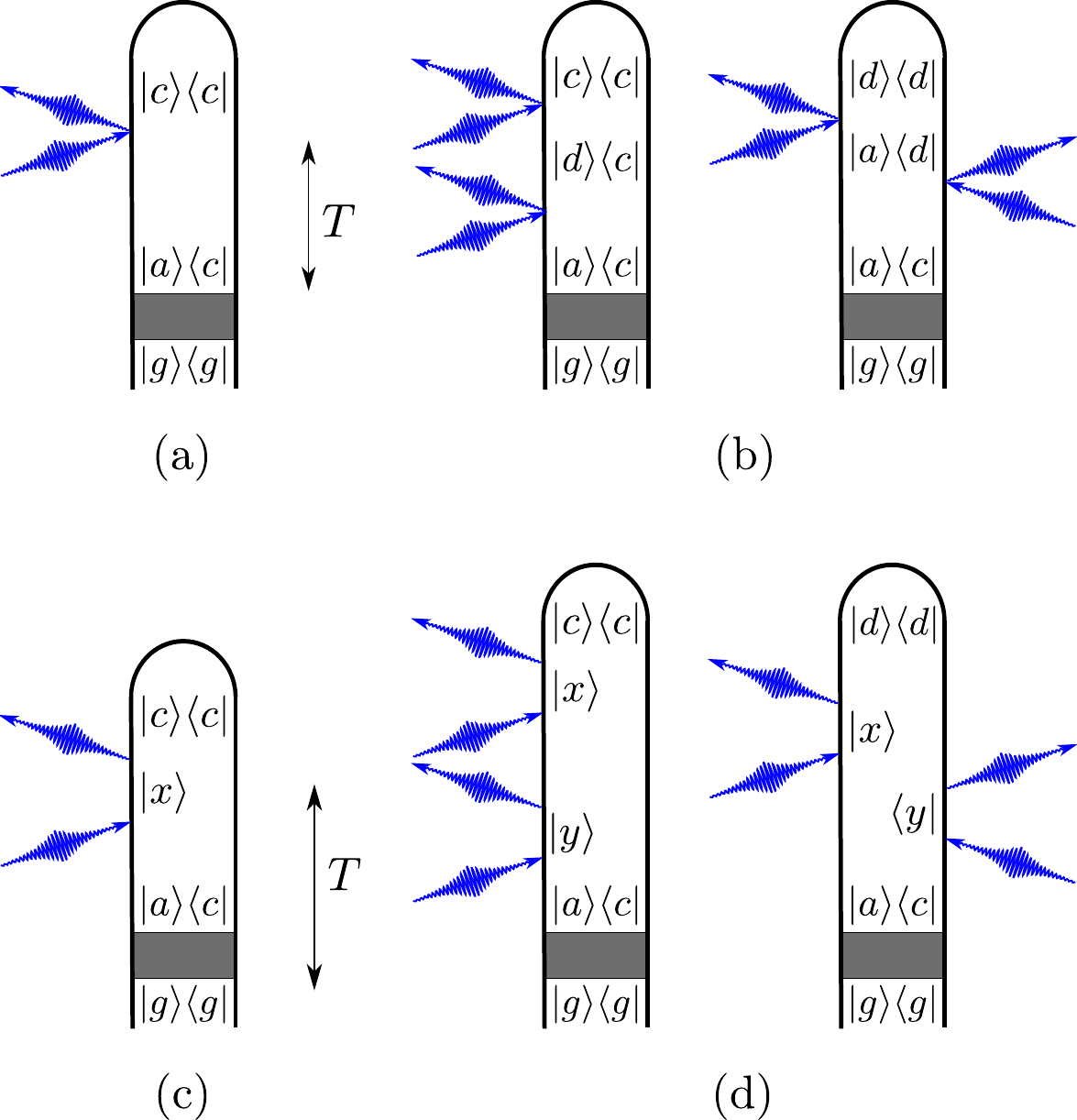}
\caption{(Color online) Loop diagrams representing the off-resonant linear (a) and quadratic (b) signals as well as the resonant linear (c) and quadratic (d) signals.  In all diagrams, the system is assumed to be prepared from the ground state by some unspecified process (depicted by the grey rectangles).  The preparation process terminates at time $\tau_0$ after which the system evolves freely until it begins interacting with the probe which is taken to be centered at time $t_0$.  The delay parameter $T=t_0-\tau_0$  is therefore shown next to the diagrams.  Note that in the off-resonant case, no core-valence coherences are created while these are created for certain times in the resonant case.  For the linear processes ((a) and (c)), $\omega_{ca}>0$ implies a red (Stokes) contribution while the reverse condition implies a blue (anti-Stokes) contribution.  For quadratic signals, this analysis holds only for one of the two relevant diagrams and so the quadratic signals are more difficult to interpret (since it depends on the state $d$). Details of the loop diagrams are given in \cite{Rah10}.}
\label{fig:diag}
\end{figure}

We will investigate stimulated X-ray Raman signals from a molecular model system with a band of valence states $a$, $c$, $d$, etc. and a core excited state band $x$, $y$ as shown in Fig. \ref{fig:scheme} (though, as mentioned in the introduction, the formalism applies as well for other different energy regimes e.g. $x$, $y$ valence excitations and $a$, $c$, $d$ vibrations). We assume that the system is prepared at time $\tau_0$ in a superposition of valence states $\psi_i$ and is monitored by interaction with a broadband probe pulse centered around time $t_0$. This is depicted diagramatically in Fig \ref{fig:diag} for both linear and quadratic signals.  Note that, although off-resonant diagrams can be deduced from the corresponding resonant diagrams, they represent prosesses governed by different interaction Hamiltonians (Eq.~(\ref{eq:Hoffres}) versus Eq.~(\ref{eq:Hintres}) for resonant interaction).  We therefore include both sets of diagrams for completeness.  A brief summary of different detection protocols for linear and quadratic signals and their most striking features in the off-resonant case are shown in Table \ref{TABLE1}. The field-matter interaction Hamiltonian for an off-resonant Raman process  in the interaction picture is given by
\begin{align}\label{eq:Hoffres}
H'(t)=\alpha^{(0)}(t)|\mathcal{E}(t)|^2,
\end{align}
where $\alpha^{(0)}$ represents the off-resonant polarizability, which is a Hermitian operator in the valence space ($\alpha^{(0)}=\alpha^\dagger+\alpha$ is real). In the following, we assume that the  electric field consists of short pulses which are temporally well-separated from the preparation process and Fourier transformable $\mathcal{E}(t)=\int_{-\infty}^{\infty}\frac{d\omega}{2\pi}\mathcal{E}(\omega)e^{-i\omega t}$. This simplifies the analysis while retaining all the essential physics. The frequency-dispersed transmission of the probe pulse (heterodyne-detected, frequency-dispersed photon-number change) is given by
\begin{align}\label{eq:Str0}
S^{(fd)}(\omega,t_0,\tau_0)=\frac{2}{\hbar}\mathcal{I}\int_{-\infty}^{\infty}dte^{i\omega(t-t_0)}\mathcal{E}^{*}(\omega)\mathcal{E}(t-t_0)\langle\langle I| \alpha^{(0)}_L(t)e^{-\frac{i}{\hbar}\int H'_-(\tau)d\tau}|\rho_i(\tau_0)\rangle\rangle,
\end{align}where $\rho_i=\vert\psi_i\rangle\langle\psi_i\vert$ is the initial density matrix (immediately following the end of the state preparation process), double brackets denote Hilbert space operators that are vectors in Liouville space, $\langle\langle I|$ represents the trace operator in Liouville space, and $\mathcal{I(R)}$ denotes the imaginary (real) part.   We adopt a superoperator notation that provides a convenient bookkeeping of time-ordered Green's functions. With any ordinary operator $A$ we associate two superoperators defined by their action on an ordinary operator $X$ as $A_L\equiv AX$  (action from the left) and $A_R\equiv XA$ (action from the right). We further define the symmetric and antisymmetric combinations $A_+=\frac{1}{2}(A_L+A_R)$, $A_-=\frac{1}{2}(A_L-A_R)$. The $H'_-$ exponential represents evolution of the matter with the probe field (since the preparatory processes have ceased) until the final interaction with the probe. We can then expand the signal perturbatively in $H'_-$ to obtain signals with the desired field-scaling (linear or quadratic in intensity are considered in this paper). 
\par
\begin{table}
  \begin{tabular}{|p{0.1\textwidth}|p{0.4\textwidth}|p{0.4\textwidth}|}
      \hline
    \multicolumn{3}{|c|}{Summary of Off-Resonant Stimulated Raman Techniques.}\\
    \hline
    \hline
    Signal & $S^{(fd)}$ & $S^{(E)}$ \\
    \hline
     $S_{LB}$ & Eq. (\ref{eq:Str2}): Oscillatory gain/loss pattern shows Stokes/anti-Stokes oscillations in $T$.  &  Eq. (\ref{eq:SE2}): No spectral resolution due to $\omega$ integration.  May visualize weak transitions due to weighting $\omega_{ac}$.  \\
    \hline
       $S_{LH}$ & Eq. (\ref{eq:Str3}): High spectral resolution compare to $S_{LB}$. Each peak oscillates at it's frequency $\omega_{ac}$ and with phase $\phi_{ac}^{\rho}$.  & Eq. (\ref{eq:SE3}): Does not carry new information compared to $S^{(fd)}$. \\
    \hline
    \hline
       $S_{QB}$ & Eq. (\ref{eq:Sqi1}): Always Stokes due to dominance of populations (which do not contribute to linear signals). & Eq. (\ref{eq:SE5}): Transition spectra can be achieved by Fourier transform over $T$. \\
    \hline
         $S_{QH}$ & Eq. (\ref{eq:Str6}): spectral and temporal resolutions are higher than $S_{LH}$ (not conjugated) but retrieval of frequencies and phases is more complicated. &  Eq. (\ref{eq:SORQHE}): with Fourier transform over $T$ can visualize weak transitions. \\
    \hline
    \hline
  \end{tabular}
  \caption{Summary of off-resonant Stimulated Raman Techniques. Note, that signals $S^{(N)}$ vanish for both linear and quadratic detection protocols, so it has been omitted.}\label{TABLE1}
\end{table}
Alternatively one can detect the total change in the photon number which is given by the zeroth spectral moment of $S^{(fd)}$
\begin{align}\label{eq:SN0}
S^{(N)}(t_0,\tau_0)&=\int_{-\infty}^{\infty}\frac{d\omega}{2\pi}S^{(fd)}(\omega,t_0,\tau_0).
\end{align}
This gives
\begin{align}
S^{(N)}(t_0,\tau_0)&=\frac{2}{\hbar}\mathcal{I}\int_{-\infty}^{\infty}dt|\mathcal{E}(t-t_0)|^2\langle\langle I| \alpha^{(0)}_L(t)e^{-\frac{i}{\hbar}\int H'_-(\tau)d\tau}|\rho_i(\tau_0)\rangle\rangle.
\end{align}
A third type of signal is given by the total change in the energy of the transmitted pulse which is the first spectral moment of $S^{(fd)}$
\begin{align}\label{eq:SE0}
S^{(E)}(t_0,\tau_0)&=\int_{-\infty}^{\infty}\frac{d\omega}{2\pi}\hbar\omega S^{(fd)}(\omega,t_0,\tau_0),
\end{align}
also given by
\begin{align}\label{eq:SE01}
S^{(E)}(t_0,\tau_0)&=-2\mathcal{R}\int_{-\infty}^{\infty}dt\dot{\mathcal{E}}^{*}(t-t_0)\mathcal{E}(t-t_0)\langle\langle I|\alpha^{(0)}_L(t)e^{-\frac{i}{\hbar}\int H'_-(\tau)d\tau}|\rho_i(\tau_0)\rangle\rangle.
\end{align}
  $S^{(fd)}$ is the most detailed measurement whereas $S^{(N)}$ and $S^{(E)}$ are its zeroth and first spectral moments. The latter two techniques are simpler to implement experimentally since they do not require frequency-selective detectors or spectrometers.  In the following we calculate the three signals (\ref{eq:Str0}) - (\ref{eq:SE01}) for a model system probed by different choices of pulse shapes and parameters.

\subsection{Off-Resonant Signals Linear in the Probe}

This technique is essentially a stimulated Raman pump-probe and is the off-resonant analogue of transient absorption. Here, the photons undergo a redistribution among field modes rather than being absorbed. Expanding Eq (\ref{eq:Str0}) to linear order in the field, we obtain
\begin{align}\label{eq:Str11}
S_L^{(fd)}(\omega,t_0,\tau_0)=\mathcal{I}2i\int_{-\infty}^{\infty}dte^{i\omega(t-t_0)}\mathcal{E}^{*}(\omega)\mathcal{E}(t-t_0)\langle\langle I| \alpha^{(0)}_L \mathcal{G}(t-\tau_0)|\rho_i\rangle\rangle,
\end{align}
where $\mathcal{G}(t)=(-i/\hbar)\theta(t)e^{-\frac{i}{\hbar}H_{0-}t}$ is the Liouville space Green's function of the molecule.  In some applications, it is necessary to carry out an exact numerical propagation of the wave function rather than using the density matrix. We can then set
\begin{align}\label{eq:Str1}
\langle\langle I| \alpha^{(0)}_L \mathcal{G}(t-\tau_0)|\rho_i\rangle\rangle=-i\hbar\langle \psi_i|G^{\dagger}(t-\tau_0) \alpha^{(0)} G(t-\tau_0)|\psi_i\rangle,
\end{align}
where $G(t)=(-i/\hbar)\theta(t)e^{-\frac{i}{\hbar}H_0t}$ is a Hilbert space Green's function. This result can be further recast in the Heisenberg picture as
\begin{align}\label{eq:Str1}
\langle\langle I| \alpha^{(0)}_L \mathcal{G}(t-\tau_0)|\rho_i\rangle\rangle=-i\hbar\langle \psi_i| \alpha^{(0)}(t-\tau_0)|\psi_i\rangle.
\end{align}

\subsubsection{Linear Broadband (LB) Probe}

We now expand the signals (\ref{eq:Str1}), (\ref{eq:SN0}), and (\ref{eq:SE0}) in molecular eigenstates . Using Fig.~\ref{fig:diag}(a), we obtain
\begin{align}\label{eq:Str2}
S_{LB}^{(fd)}(\omega,T)=-\frac{2}{\hbar}\sum_{a,c}|\mathcal{E}(\omega)||\mathcal{E}(\omega-\omega_{ac})|\alpha^{(0)}_{ca}|\rho_{ac}|\sin\phi_{ac}(T),
\end{align}
\begin{align}\label{eq:SN2}
S_{LB}^{(N)}(T)=0,
\end{align}
\begin{align}\label{eq:SE2}
S_{LB}^{(E)}(T)=-\frac{1}{\hbar}\sum_{a,c}\int_{-\infty}^{\infty}\frac{d\omega}{2\pi}|\mathcal{E}(\omega)||\mathcal{E}(\omega-\omega_{ac})|\alpha^{(0)}_{ca}|\rho_{ac}|\hbar\omega_{ac}\sin\phi_{ac}(T).
\end{align}
Here, $\rho_{ac}=|\rho_{ac}|e^{i\phi_{ac}^{\rho}}$ is the initial density matrix, the phase is given by $\phi_{ac}(T)=\omega_{ac}T-\phi_{ac}^{\rho}$, and we neglect the linewidth for the valence states. In Eq.~(\ref{eq:Str2}), terms in which  $\omega_{ac}>0$ ($\omega_{ac}<0$) lead to broad peaks below (above) the central pulse frequency. We denote these as``red" and ``blue" peaks respectively.  Each pair of states therefore generates a complementary pair of red and blue contributions.  The two  oscillate in $T$ with a $\pi$ phase shift (as sine is an odd function of its argument).  The frequency-dispersed signal therefore oscillates between Stokes (positive red contributions and negative blue) and anti-Stokes (negative red contributions and positive blue) processes as shown in Fig.~\ref{fig:SLBQB}(a1). 
\par
Equation (\ref{eq:SE2}) is plotted in Fig.~\ref{fig:SLBQB}(b1) but, aside from indicating that some nontrivial phase is involved (since the signal doesn't vanish for $T\to 0$), not much information is directly apparent from the time-domain measurement.  Taking the Fourier transform $\int dTe^{i\Omega T}S(T)=S(\Omega)$ (Fig.~\ref{fig:SLBQB}(c1)) reveals peaks at each transition energy $\omega_{ac}$ whose heights are given by the factor $\omega_{ac}\alpha_{ca}\vert\rho_{ac}\vert$.  The ratios of these peak heights thus give information on the polarizabilities or the magnitudes of the initial coherences. Finally, it is worth noting that $\phi_{aa}(T)=0,\forall T$ so that populations do not contribute to the off-resonant linear broadband signal.  \textit{These signals therefore provide a background-free detection of the electronic coherences. }

\subsubsection{Linear hybrid (LH) probe}

\begin{samepage}
A hybrid probe is a shaped pulse consisting of a broadband attosecond pulse $\mathcal{E}(\omega)=|\mathcal{E}_0(\omega)|e^{i\phi_0}$ and a narrowband  femtosecond pulse $\mathcal{E}_1(t)=|\mathcal{E}_1|e^{-i\omega_1t+i\phi_1}$  centered at frequency $\omega_1$. In this case, Eq.~(\ref{eq:Str11}) yields
\begin{align}\label{eq:Str3}
&S_{LH}^{(fd)}(\omega-\omega_1,\omega_1,T)=-\frac{4\pi}{\hbar}|\mathcal{E}_0(\omega)||\mathcal{E}_1|\sum_{a,c}\alpha^{(0)}_{ca}|\rho_{ac}|\sin\phi_{ac}^1(T)\delta(\omega-\omega_1-\omega_{ac}),
\end{align}
where $\phi_{ac}^1(T)=\omega_{ac}T-\phi_{ac}^{\rho}+\phi_0-\phi_1$. Note that if the broadband and the narrowband components have the same phase, $\phi_0=\phi_1$, then $\phi_{ac}^1(T)=\phi_{ac}(T)$. Due to the dependence on the relative phase of the pulses $\phi_{0}-\phi_1$, observation of the linear hybrid signals requires phase-control (averaging over random $\phi_{0}$, $\phi_1$ causes the signal to vanish). Eq.~(\ref{eq:Str3}) yields clearly resolved Raman resonances unlike the $LB$ case where the red ($\omega_{ac}>0$) and blue ($\omega_{ac}<0$) components only enter through the pulse envelope. This signal is depicted for $\omega_1=\omega_0$ in Fig.~\ref{fig:SLHQH}(a1), shows sharp peaks at each transition frequency $\omega_{ac}$.  Of particular significance is the ability to extract the phase $\phi^{\rho}_{ac}$ from the oscillations of the separate peaks. This information is not available in the $LB$ signal.
\end{samepage}
\par
The photon number signal is slightly more complex: because the hybrid pulse contains both broad and narrowband components we need to take into account the contribution to the signal where the last interaction is with narrowband component. The total change in photon number vanishes, as in the broadband case,
\begin{align}\label{eq:SN3}
S_{LH}^{(N)}(\omega_1,T)=0.
\end{align}
The transmitted energy change of the shaped pulse similarly contains both narrowband and broadband components and the total energy change of the shaped pulse is given by
\begin{align}\label{eq:SE3}
S_{LH}^{(E)}(\omega_1,T)=-\frac{4\pi}{\hbar}\sum_{a,c}|\mathcal{E}_1||\mathcal{E}_0(\omega_1-\omega_{ac})|\alpha^{(0)}_{ca}|\rho_{ac}|\hbar\omega_{ac}\sin\phi_{ac}^1(T).
\end{align}
Note the similarity between this and Eq.~(\ref{eq:Str2}) (they only differ by the factor $\omega_{ac}\mathcal{E}(\omega)$).  The $\omega$ integration erodes the sharp resolution afforded by the narrowband pulse, leaving virtually the same result as the $S^{(fd)}_{LB}$ signal (Eq.~\ref{eq:Str2}). The $S^{(E)}_{LH}$ signal is shown in Fig.~\ref{fig:SLHQH}(b1).  Note that the presence of the $\omega_{ac}$ factor which reverses the sign under $a\to c$ relative to Eq.~(\ref{eq:Str2}) rendering both red and blue contributions the same sign for a given $T$. The sign of $S^{(E)}_{LH}(T)$ then indicates whether the process is Stokes ($S^{(E)}(T)<0$) or anti-Stokes ($S^{(E)}(T)>0$). Energy conservation implies that the pulse energy change and the molecular energy change must have equal magnitude and opposite sign. In particular, the diagram in Fig.~\ref{fig:diag}(a) assumes that, after preparation, the molecule is in the superposition of states $a$ and $c$ and the final state of the molecule after interacting with transmitted pulse is a population $cc$. Similarly using the permutation $a\leftrightarrow c$ one can end up in the final state $aa$. If the energy of state $a$ is higher (lower) than that of $c$ this is a Stokes (anti-Stokes) process. The molecular energy change is given by
\begin{align}
S^{(E)}(t_0,\tau_0)=-\sum_{a,c}\hbar\omega_{ac}(\rho_{aa}(t_0,\tau_0)-\rho_{cc}(t_0,\tau_0)).
\end{align}
This result holds for an arbitrarily prepared molecular state prior to the probe.

\subsection{Off-Resonant Signals Quadratic in the Probe}

The quadratic signal (i.e., second-order in probe intensity) is obtained by expanding the exponent in Eqs.~(\ref{eq:Str0})-(\ref{eq:SE0}) to first-order. The signal can be  read off the two diagrams shown in Fig.~\ref{fig:diag}(b), yielding
\begin{align}\label{eq:Str411}
S_{Q}^{(fd)}(\omega,t_0,\tau_0)&=\mathcal{I}2i\int_{-\infty}^{\infty}dt\int_{-\infty}^tdt'e^{i\omega(t-t_0)}\mathcal{E}^{*}(\omega)\mathcal{E}(t-t_0)|\mathcal{E}(t'-t_0)|^2\notag\\
&\times\langle\langle I|\alpha^{(0)}_L \mathcal{G}(t-t')\alpha^{(0)}_- \mathcal{G}(t'-\tau_0)|\rho_i\rangle\rangle,
\end{align}
\begin{align}\label{eq:SN411}
S_{Q}^{(N)}(t_0,\tau_0)&=\mathcal{I}2i\int_{-\infty}^{\infty}dt\int_{-\infty}^tdt'|\mathcal{E}(t-t_0)|^2|\mathcal{E}(t'-t_0)|^2\notag\\
&\times\langle\langle I|\alpha^{(0)}_L \mathcal{G}(t-t')\alpha^{(0)}_- \mathcal{G}(t'-\tau_0)|\rho_i\rangle\rangle,
\end{align}
\begin{align}\label{eq:SE411}
S_{Q}^{(E)}(t_0,\tau_0)&=2\hbar\mathcal{I}\int_{-\infty}^{\infty}dt\int_{-\infty}^tdt'\dot{\mathcal{E}}^{*}(t-t_0)\mathcal{E}(t-t_0)|\mathcal{E}(t'-t_0)|^2\notag\\
&\times\langle\langle I|\alpha^{(0)}_L \mathcal{G}(t-t')\alpha^{(0)}_- \mathcal{G}(t'-\tau_0)|\rho_i\rangle\rangle,
\end{align}
The corresponding Hilbert space expressions which are suitable for wavefunction-based simulations are
\begin{align}\label{eq:Str4100}
\langle\langle I|\alpha^{(0)}_L \mathcal{G}(t-t')\alpha^{(0)}_- \mathcal{G}(t'-\tau_0)|\rho_i\rangle\rangle=-i\hbar[&\langle \psi_i|G^{\dagger}(t-\tau_0) \alpha^{(0)} G(t-t')\alpha^{(0)} G(t'-\tau_0)|\psi_i\rangle\notag\\
-&\langle \psi_i|G^{\dagger}(t'-\tau_0) \alpha^{(0)} G^{\dagger}(t-t')\alpha^{(0)} G(t-\tau_0)|\psi_i\rangle]
\end{align}
Eq. (\ref{eq:Str4100}) may be alternatively recast in the Heisenberg representation
\begin{align}\label{eq:Str41100}
\langle\langle I|\alpha^{(0)}_L \mathcal{G}(t-t')\alpha^{(0)}_- \mathcal{G}(t'-\tau_0)|\rho_i\rangle\rangle=-i\hbar[&\langle \psi_i| \alpha^{(0)}(t-\tau_0) \alpha^{(0)}(t'-\tau_0)|\psi_i\rangle\notag\\
-&\langle \psi_i| \alpha^{(0)}(t'-\tau_0) \alpha^{(0)}(t-\tau_0)|\psi_i\rangle].
\end{align}

\subsubsection{Quadratic Broadband (QB) Probe}

The frequency-dispersed quadratic signal (\ref{eq:Str411}) for a broadband probe expanded in eigenstates reads 
\begin{align}\label{eq:Sqi1}
S_{QB}^{(fd)}(\omega,T)&=-\frac{2}{\hbar^2}\int\frac{d\omega_1}{2\pi}|\mathcal{E}(\omega)||\mathcal{E}(\omega_1)|\sum_{a,c,d}\alpha^{(0)}_{cd}\alpha^{(0)}_{da}|\rho_{ac}|\cos\phi_{ac}(T)\notag\\
&\times[|\mathcal{E}(\omega-\omega_{dc})|\mathcal{E}(\omega_1+\omega_{da})|-|\mathcal{E}(\omega+\omega_{da})||\mathcal{E}(\omega_1-\omega_{dc})|],
\end{align}
where the first (second) term in the square brackets corresponds to the left (right) diagrams in Fig.~\ref{fig:diag}(b).
Eq.~(\ref{eq:Sqi1}) has a more complex dependence on the Raman shift $\omega_{ac}$ than the linear signal (\ref{eq:Str2}). The field envelopes are now shifted by the electronic transition frequencies $\omega_{ad}$ and $\omega_{dc}$ which yields difference $\omega_{dc}-\omega_{da}=\omega_{ac}$. The quadratic signal oscillates with a phase that depends on states $d$  other than $a$ and $c$ that create the resonance and involves the phases of the polarizability $\alpha^{(0)}$.

Comparing this to the corresponding linear signal (Eq.~(\ref{eq:Str2})) shows that a $\pi/2$ phase change.  As a result, the fact that $\phi_{aa}(T)=0$ no longer eliminates contributions from populations and they form a time-dependent background to all quadratic signals.  Because the initial density matrix is assumed to be perturbative (so that the ground state dominates the populations) the contribution due to populations is primarily Stokes overall.  Since the contribution due to the populations is Stokes and the oscillating coherences are too weak to overcome this, the overall process is Stokes at all times $T$ (as illustrated in Fig.~\ref{fig:SLBQB})(a2).

The integrated photon number vanishes $S_{QB}^{(N)}(T)=0$. The energy change signal (\ref{eq:SE0})  is given by
\begin{align}\label{eq:SE5}
S_{QB}^{(E)}(T)&=-\frac{1}{\hbar^2}\int\frac{d\omega}{2\pi}\int\frac{d\omega_1}{2\pi}|\mathcal{E}(\omega)||\mathcal{E}(\omega_1)|\sum_{a,c,d}|\mathcal{E}(\omega+\omega_{da})||\mathcal{E}(\omega_1-\omega_{dc})|\notag\\
&\times(\hbar\omega_{da}+\hbar\omega_{dc})|\rho_{ac}|\alpha_{cd}^{(0)}\alpha_{da}^{(0)}\cos\phi_{ac}(T).
\end{align}
Since the populations contribute a static off-set, there is now a strong zero-frequency peak in the Fourier transform of Eq.~(\ref{eq:SE5}).  Additionally, the peak heights are no longer as simply related to the $\alpha_{ac}$, $\rho_{ac}$ as in the linear case.  In particular, since the peak heights are determined by a free summation over the intermediate state $d$, the terms in this sum can interfere constructively or destructively leading to enhanced or suppressed peaks (note that the $\omega_{ac}=4\text{~eV}$ peak is suppressed in Fig.~\ref{fig:SLBQB})(c2).

\subsubsection{Quadratic Hybrid (QH) Probe}

By expanding the quadratic frequency-dispersed transmission signal (\ref{eq:Str411}) in eigenstates we obtain
\begin{align}\label{eq:Str6}
S_{QH}^{(fd)}&(\omega-\omega_1,\omega_1,T)=-\frac{4\pi}{\hbar^2}|\mathcal{E}_0(\omega)||\mathcal{E}_1|^2\sum_{a,c,d}\alpha^{(0)}_{da}\alpha^{(0)}_{cd}|\rho_{ac}|\cos\phi_{ac}(T)\notag\\
&\times\left[|\mathcal{E}_0(\omega_1+\omega_{da})|\delta(\omega-\omega_1-\omega_{dc})-|\mathcal{E}_0(\omega_1-\omega_{dc})|\delta(\omega-\omega_1+\omega_{da})\right].
\end{align}
As in the linear case, the narrowband pulse allows us to clearly resolve the transition peaks.  Unlike the linear signal (\ref{eq:Str3}), the quadratic hybrid signal is independent of the phases of the narrowband and broadband pulses $\phi_1$ and $\phi_0$ and is therefore observable without phase-controlled pulses.  Just as in the broadband case, the signal is sensitive to populations which contribute a static Stokes spectrum.  In Fig.~\ref{fig:SLHQH}(a2), we plot this signal for $\omega_1=\omega_0$ as well as separate plots for the contributions due to populations and coherences.
\par
To calculate the integrated signals we must also include contributions from diagrams whereby the last interaction is with narrowband pulse $\mathcal{E}_1$. The total photon number change then vanishes $S_{QH}^{(N)}(\omega_1,T)=0$. The corresponding pulse energy change which includes both broadband and narrowband components is
\begin{align}\label{eq:SORQHE}
S_{QH}^{(E)}(\omega_1,T)&=-\frac{2}{\hbar^2}|\mathcal{E}_1|^2\sum_{a,c,d}\alpha^{(0)}_{cd}\alpha^{(0)}_{da}|\rho_{ac}|\cos\phi_{ac}(T)\notag\\
&\times [\hbar\omega_{dc}|\mathcal{E}_0(\omega_1+\omega_{dc})||\mathcal{E}_0(\omega_1+\omega_{da})|+\hbar\omega_{da}|\mathcal{E}_0(\omega_1-\omega_{da})||\mathcal{E}_0(\omega_1-\omega_{dc})|].
\end{align}
Note that in the quadratic case the energy flux involves also another state $d$ in addition to states $a$ and $c$, so the different fluxes corresponding to all relevant pairs of states should be added to get the overall energy change of the pulse. 

\subsection{Discussion of Off-Resonant Signals}

\begin{figure}
\centering
\includegraphics[width=.85\textwidth]{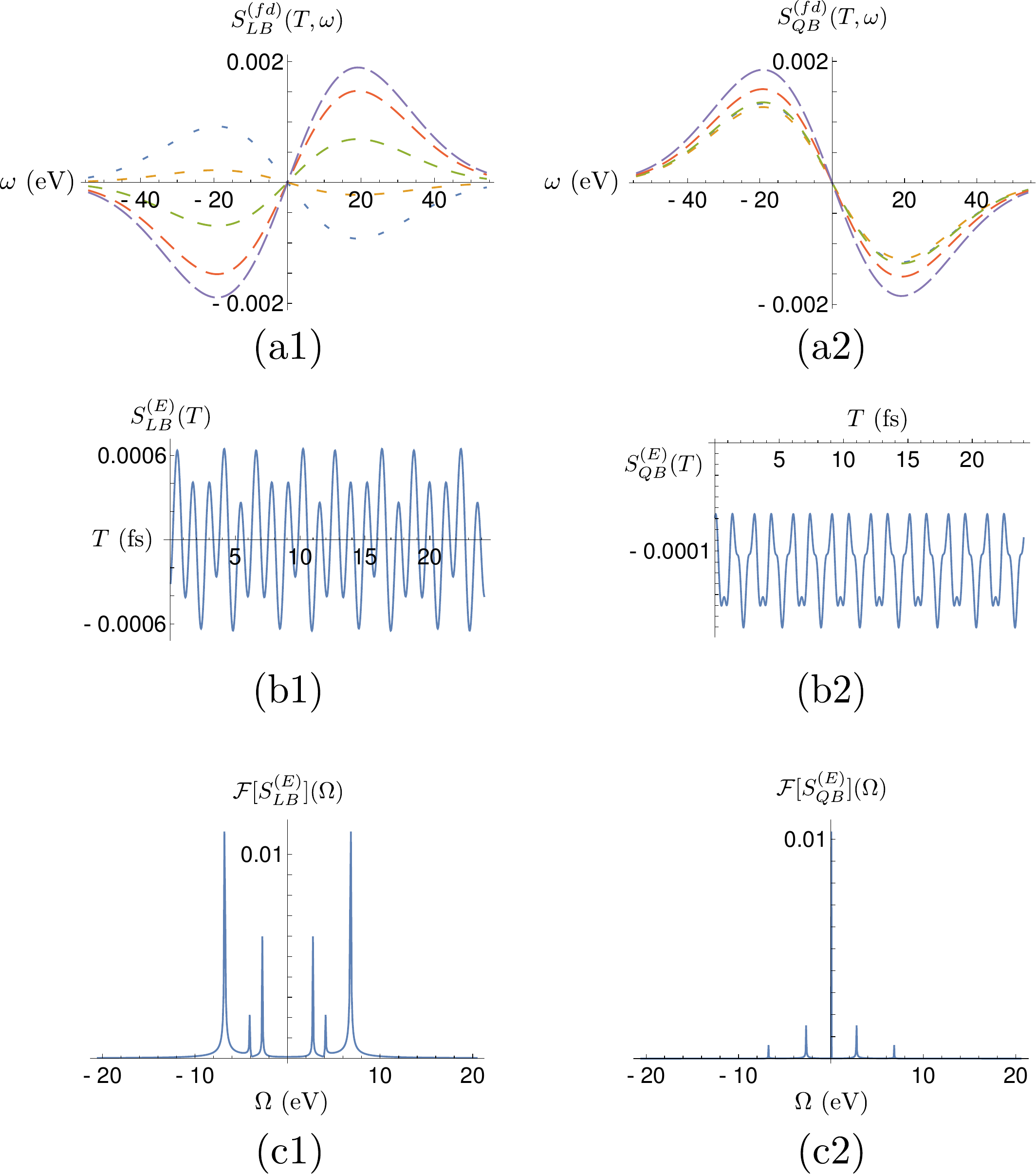}
\caption{(Color online) (a1) and (a2): Off-resonant linear and quadratic frequency-dispersed signals (Eqs. (\ref{eq:Str2}) and (\ref{eq:Sqi1})) at various times $T$ after state preparation ($T$ advances in units of $\sim$60 attoseconds as the dashes lengthen).   (b1) and (b2): Off-resonant linear and quadratic broadband energy signal as a function delay time $T$ (Eqs. (\ref{eq:SE2}) and (\ref{eq:SE5})).  Note that the linear signal oscillates about zero but the quadratic signal has a static offset corresponding to the contribution from populations.  (c1) and (c2): Fourier transforms of (b1) and (b2).  In the linear case, peaks corresponding to all $\omega_{ca}$ coherences are visible.  In the quadratic case, there is a large central peak at zero corresponding to the populations.  In the quadratic signal, the states $c-a$ are coupled indirectly through a third state $d$ and the summation over intermediate states can suppress or enhance the Raman peak magnitudes relative to their linear proportions.}
\label{fig:SLBQB}
\end{figure}
\begin{figure}
\centering
\includegraphics[width=.85\textwidth]{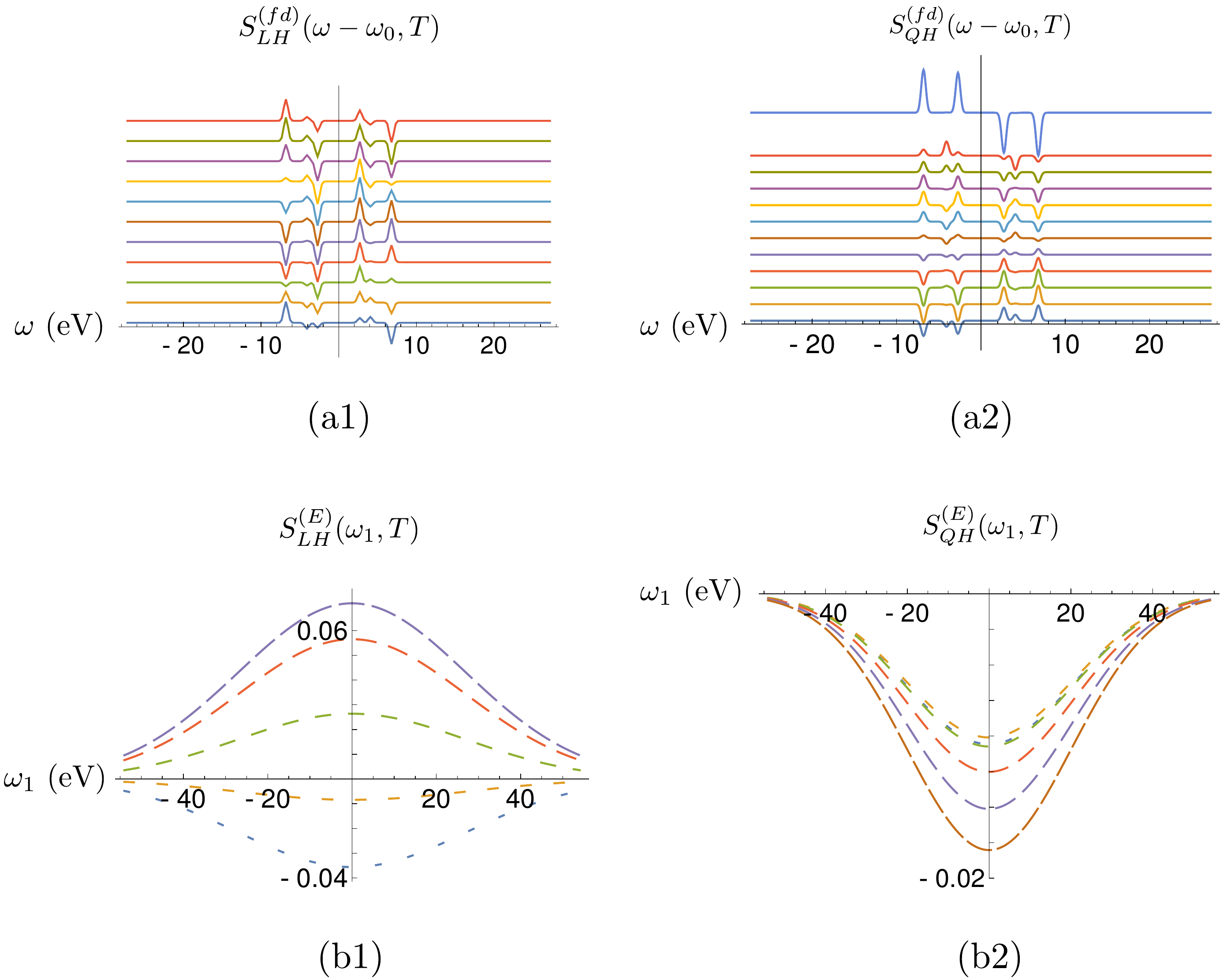}
\caption{(Color online) (a1) and (a2):Off-resonant linear and quadratic hybrid frequency-dispersed signals (Eqs. (\ref{eq:Str3}) and (\ref{eq:Str6})) at various times $T$ after state preparation ($T$ advances in units of $\sim$60 attoseconds as one goes up the vertical axis).  The quadratic signal is split into the static contribution due to populations (a2 top spectrum) and the time-dependent contribution due to coherences (a2 lower spectra).  In contrast to the broadband signals, the narrowband pulse allows us to resolve the individual transition peaks.  (b1) and (b2): Off-resonant linear and quadratic hybrid energy signals (Eqs. (\ref{eq:SE3}) and (\ref{eq:SORQHE})) as a function of narrowband frequency $\omega_1$ and delay time $T$ ($T$ advances in units of $\sim$60 attoseconds as the dashes lengthen).  The broadband detection renders individual transition peaks unobservable and the result is similar to the broadband case (Eq.~(\ref{eq:SE5})) but symmetric (rather than anti-symmetric) about $\omega=\omega_0$.  Note that $S_{QH}^{(E)}(T)<0$ revealing that the quadratic process is always Stokes while the linear process oscillates between Stokes and anti-Stokes.}
\label{fig:SLHQH}
\end{figure}

In the previous section we considered a variety of heterodyne-detected signals in the off-resonant regime. Here, we present these signals calculated for a simple model consisting of a ground state and two valence excitations at $\varepsilon_{a(c)}\in\left\{ {0,2.7,6.8}\right\}(\text{eV})$.  The polarizabilities $\alpha_{ac}$ are symmetric but otherwise random numbers of $\mathcal{O}(.1-1)$ and the $\rho_{ac}$ are random but taken from a pure perturbative state (i.e., a state $\vert\psi\rangle$ in which the ground state amplitude is near unity in magnitude).  
\par
Figure \ref{fig:SLBQB} shows the linear and quadratic off-resonant broadband signals side-by-side.  We do not plot $S^{(N)}$ since it vanishes in the off-resonant regime.  For the frequency-dispersed signal (Eqs.~(\ref{eq:Str2}), (\ref{eq:Str3}), (\ref{eq:Sqi1}), and (\ref{eq:Str6})), the broadband detection makes discerning individual transitions impossible but nicely illustrates the Stokes/anti-Stokes oscillations in time. In the linear case, the red contributions are positive and the blue are negative at $T=0$ so the process is Stokes at this time and oscillates with $T$.  The quadratic detection is always Stokes due to the effect of populations.  The time-domain energy signal (Eqs.~(\ref{eq:SE2}), (\ref{eq:SE3}), (\ref{eq:SE5}), and (\ref{eq:SORQHE})) reveals again that the quadratic process is always Stokes while the linear process oscillates between Stokes and anti-Stokes.  Transforming these signals reveals the $\omega_{ca}$ transitions (though the quadratic possesses a large $\Omega=0$ peak due to populations that is of course missing from the linear signal).
\par
For comparison, the linear and quadratic hybrid signals are shown in figure \ref{fig:SLHQH}.  The spectral resolution of the narrowband pulse gives sharp peaks at the $\omega-\omega_1=\omega_{ac}$.  In the linear case, each $\omega_{ac}$ peak is due to a term that oscillates with this same frequency.  However, in the quadratic case, each oscillating term contains peaks at all different $\omega_{da}$, $\omega_{dc}$ (as per Eq.~(\ref{eq:Str6})) and the phases of the peak oscillations do not as directly reveal the phases $\phi_{ac}^\rho$.  Note that the free summation in the quadratic signal implies that peaks will be visible even for states not initially occupied (i.e.~for which $\rho_{ac}=0$) while the peaks observable in the linear signal are restricted to those initially occupied. 
\par
In summary,  $S^{(fd)}$ shows an oscillatory pattern of gain and loss features in the red and blue regimes that depend on the initial phase of $\rho_{ac}$. This information is integrated out in $S^{(E)}$ where the entire probe pulse envelope exhibits Stokes (loss) or anti-Stokes (gain) shifts periodically \cite{ideguchi2013coherent}. The broadband signals alone do not possess a sufficient resolution to directly observe the transition spectra and only allow us to access it through the Fourier transforms of the energy signals $S^{(E)}(T)$.  Utilizing a hybrid broad-narrow pulse combined with frequency-dispersed detection allows high spectral and temporal resolution thus permitting spectral snapshots to be taken that clearly resolve all transitions and therefore permit extraction of the phases from the oscillation patterns.  In the linear case, this requires control of the relative pulse phases (the phases cancel in the quadratic case).  The quadratic and linear signals generally carry the same information about coherences but this is accessed without the background due to populations in the case of linear signals.  One important caveat to this is that the magnitudes of the various transition peaks can be enhanced or suppressed by the interference of different pathways (characterized by the intermediate state $d$).  Since each oscillation term contributes to multiple peaks through this intermediate state, a given peak $\omega_{ac}$ will not oscillate at the corresponding phase $\phi^{\rho}_{ac}$, making retrieval of the wavepacket's phase more difficult in the quadratic case. 
 
Equations (\ref{eq:Str1}) and (\ref{eq:Str4100}) can be used for the numerical simulation of signals from systems undergoing complex dynamics. The Liouville space form can describe the reduced density matrix by using the SLE. The Hilbert space form is suitable for direct numerical simulation of the wavefunction that includes all degrees of freedom. It can apply e.g. to nonadiabatic dynamics at conical intersections. The eigenstate expansions (Eqs.~(\ref{eq:Str2})-(\ref{eq:SE3}) and (\ref{eq:Sqi1})-(\ref{eq:SORQHE})) may be used when the probe is impulsive so that the eigenstates do not vary during the probing process.  Appendix \ref{app:MultiRaman} extends this treatment of off-resonant signals to a two-pulse time-domain experiment and considers the preparation by an actinic pulse.

\section{Resonant Stimulated Raman Techniques}\label{sec:Res}

The off-resonant techniques considered in section \ref{sec:OffRes} conserve the number of probe photons.  Photon energy gets redistributed among the broadband modes: $S^{(N)}$ vanishes while $S^{(E)}$ is finite.  When the X-ray pulses are resonant with core transitions, true photon absorption can take place.  This renders $S^{(N)}$ finite as well revealing new matter information. Below we discuss both linear and quadratic resonant signals.

\subsection{Linear Probe}

The signal linear in the probe intensity is given by the diagram shown in Fig.~\ref{fig:diag}(c). Due to resonant excitation, the field-matter interaction Hamiltonian may no longer be recast using a frequency-independent, off-resonant polarizability but should rather be described by a dipole interaction Hamiltonian in the rotating wave approximation (RWA)
\begin{align}\label{eq:Hintres}
H_-'(t)=V(t)E^{\dagger}(t)+V^{\dagger}(t)E(t).
\end{align}
 The frequency-dispersed signal can be read off the diagram
 \begin{align}\label{eq:Slitr1}
S_{L}^{(fd)}(\omega,t_0,\tau_0)=2\mathcal{I}i\mathcal{E}^{*}(\omega)\int_{-\infty}^{\infty}dte^{i\omega(t-t_0)}\int_{-\infty}^tdt'\mathcal{E}(t'-t_0)\langle\langle I|V_L\mathcal{G}(t-t')V_L^{\dagger}\mathcal{G}(t'-\tau_0)|\rho_i\rangle\rangle.
\end{align}
The integrated photon number is given by
\begin{align}\label{eq:SliN11}
S_{L}^{(N)}(t_0,\tau_0)=2\hbar\mathcal{I}\int_{-\infty}^{\infty}dt\int_{-\infty}^tdt'\mathcal{E}^{*}(t-t_0)\mathcal{E}(t'-t_0)\langle\langle I|V_L\mathcal{G}(t-t')V_L^{\dagger}\mathcal{G}(t'-\tau_0)|\rho_i\rangle\rangle.
\end{align}
The pulse energy change is given by
\begin{align}\label{eq:SliE11}
S_{L}^{(E)}(t_0,\tau_0)=2\hbar^2\mathcal{R}\int_{-\infty}^{\infty}dt\int_{-\infty}^tdt'\frac{d\mathcal{E}^{*}(t-t_0)}{dt_0}\mathcal{E}(t'-t_0)\langle\langle I|V_L\mathcal{G}(t-t')V_L^{\dagger}\mathcal{G}(t'-\tau_0)|\rho_i\rangle\rangle.
\end{align}
The signals  (\ref{eq:Slitr1})  - (\ref{eq:SliE11}) can be alternatively recast in the Hilbert space form suitable for numerical propagation by setting
\begin{align}
\langle\langle I|V_L\mathcal{G}(t-t')V_L^{\dagger}\mathcal{G}(t'-\tau_0)|\rho_i\rangle\rangle=-i\hbar\langle \psi_i|G^{\dagger}(t-\tau_0)VG(t-t')V^{\dagger}G(t'-\tau_0)|\psi_i\rangle
\end{align}
These signals are often referred to as ``transient absorption".  

\subsubsection{Linear Broadband (LB) probe}

Expanding the frequency-dispersed transmission (\ref{eq:Slitr1}) as a sum over states and evaluating time integrals we obtain 
\begin{align}\label{eq:Slfdbb1}
S_{LB}^{(fd)}(\omega,T)=-\frac{2}{\hbar^2}\sum_{a,c}|\rho_{ac}|\left(\alpha_{ca}^{(\mathcal{E}\mathcal{E})'}(\omega)\sin\phi_{ac}(T)-\alpha_{ca}^{(\mathcal{E}\mathcal{E})''}(\omega)\cos\phi_{ac}(T)\right),
\end{align}
where $\Phi_{ac}(T)=\omega_{ac}T-\phi_{ac}^{\rho}$, $\alpha'$ and $\alpha''$  represent the real and imaginary parts of the polarizability 
\begin{align}\label{eq:def1}
\alpha_{ca}^{(jk)}(\omega)=\sum_x\frac{\tilde{\mathcal{E}}_j^{*}(\omega)\tilde{\mathcal{E}}_k(\omega+\omega_j-\omega_k+\omega_{ca})(\mathbf{e}_k\cdot V_{cx})(\mathbf{e}_j\cdot V_{xc})}{\omega+\omega_j-\omega_{xa}+i\Gamma_x},
\end{align}
where $\omega_m$, $m=j,k$ is the central frequency of the pulse given by $\mathcal{E}_m(t)=\int_{-\infty}^{\infty}\frac{d\omega}{2\pi}\tilde{\mathcal{E}}_m(\omega)e^{i(\omega+\omega_m)t}$, where we redefined the frequency domain amplitude $\tilde{\mathcal{E}}_\alpha(\omega)$ to be centered at zero frequency (for a Gaussian pulse with bandwidth $\sigma_m$, $\tilde{\mathcal{E}}_m(\omega)=\frac{1}{\sqrt{2\pi}\sigma_m}e^{-\frac{\omega^2}{2\sigma_m^2}}$), $\Gamma_e$ is the inverse excited state lifetime, which is assumed to be shorter than states $a$ and $c$. Thus, the polarizability implicitly depends on the pulse parameters, such as its central frequency. For $j=k$, Eq.~(\ref{eq:def1}) reduces to the commonly used polarizability (see Eq.~(5) of \cite{biggs}). Note that in the off-resonant case the polarizability matrix elements are real and the second term in Eq. (\ref{eq:Slfdbb1}) vanishes and the Stokes and anti-Stokes components oscillate with opposite phase as seen in Section II. The integrated photon number (\ref{eq:SliN11}) is
\begin{align}\label{eq:SlN20}
S_{LB}^{(N)}(T)=\frac{2}{\hbar^2}\sum_{a,c}|\rho_{ac}|\alpha_{ca}^{(\mathcal{E}\mathcal{E})''}\cos\phi_{ac}(T),
\end{align}
where the integrated polarizability is given by 
\begin{align}
\alpha_{ca}^{(jk)}=\int_{-\infty}^{\infty}\frac{d\omega}{2\pi}\alpha_{ca}^{(jk)}(\omega)
\end{align}
and we used the symmetry $\alpha_{\alpha\beta}^{(\mathcal{E}\mathcal{E})}=\alpha_{\beta\alpha}^{(\mathcal{E}\mathcal{E})}$. We first note that the signal is given by the imaginary part of the polarizability. Therefore, when the polarizability is real (as in the off-resonant case) the signal vanishes, which is consistent with our earlier result of Eq. (\ref{eq:SN2}).  Again the Stokes ($\omega_{ac}>0$) and anti-Stokes ($\omega_{ac}<0$) components oscillate with an opposite phase. The energy change of the pulse signal (\ref{eq:SliE11}) is given by
\begin{align}\label{eq:SEN20}
S_{LB}^{(E)}(T)=-\frac{1}{\hbar^2}\sum_{a,c}|\rho_{ac}|\left[\hbar\omega_{ac}\alpha_{ca}^{(\mathcal{E}\mathcal{E})'}\sin\phi_{ac}(T)-(\beta_{ca}^{(\mathcal{E}\mathcal{E})''}+\beta_{ac}^{(\mathcal{E}\mathcal{E})''})\cos\phi_{ac}(T)\right],
\end{align}
where the tensor $\beta$ represents the first moment of the polarizability:
\begin{align}
\beta_{ca}^{(jk)}=\int_{-\infty}^{\infty}\frac{d\omega}{2\pi}\hbar(\omega+\omega_{j})\alpha_{ca}^{(jk)}(\omega).
\end{align}
Unlike the photon number signal (\ref{eq:SlN20}), the energy change (\ref{eq:SEN20}) involves both the real part of the polarizability tensor and the imaginary part of its first moment. Eq. (\ref{eq:SEN20}) reduces to Eq. (\ref{eq:SE2}) in the off-resonant case, where the polarizability is real and the second term vanishes.

\subsubsection{Linear Hybrid (LH) probe}

The frequency-dispersed transmission of the broadband pulse $\mathcal{E}(\omega)$ (\ref{eq:Slitr1}) in the presence of the narrowband pulse $\mathcal{E}_1$ (hybrid probe) is given by
\begin{align}\label{eq:Slfdsh1}
S_{LH}^{(fd)}(\omega-\omega_1,\omega_1,T)=-\frac{2}{\hbar^2}\sum_{a,c}|\rho_{ac}|\left(\alpha_{ca}^{(01)'}(\omega)\sin\phi_{ac}(T)-\alpha_{ca}^{(01)''}(\omega)\cos\phi_{ac}(T)\right),
\end{align}
where the $\omega_1$ dependence  is now implicitly included in $\alpha_{ca}^{(01)}$ by setting $\tilde{\mathcal{E}}_1(\omega)=\tilde{\mathcal{E}}_1\delta(\omega)$.
The integrated photon number (\ref{eq:SliN11}) for a shaped pulse has to be calculated differently than the broadband case. In particular, one has to take into account the photon number change in both the broadband and narrowband pulses. This yields
\begin{align}\label{eq:SlN3}
S_{LH}^{(N)}(\omega_1,T)=-\frac{2}{\hbar^2}\sum_{a,c}|\rho_{ac}|\left([\alpha_{ca}^{(01)'}+\alpha_{ca}^{(10)'}]\sin\phi_{ac}(T)-[\alpha_{ca}^{(01)''}+\alpha_{ca}^{(10)''}]\cos\phi_{ac}(T)\right),
\end{align}
where the first (second) terms in each square bracket represents the last interaction with the broadband (narrowband) pulse. For the hybrid pulse, both $\alpha'$ and $\alpha''$ contribute to the signal, thus providing additional molecular information than the simple broadband pulse. Furthermore, the form of the anti-Stokes and Stokes polarizabilities suggests that the signal depends exclusively on the phase difference between both broadband and narrowband fields $\phi_0-\phi_1$ which provides an additional control knob. Note, that in general the hybrid polarizabilities $\alpha_{xy}^{(10)}$ are not symmetric under permutation of their indices. However in the limit when $\phi_0=\phi_1$, we have $\alpha_{xy}^{(10)}=\alpha_{yx}^{(01)}$ and Eq.~(\ref{eq:SlN3}) yields
\begin{align}\label{eq:SlN301}
S_{LH}^{(N)}(\omega_1,T)=\frac{2}{\hbar^2}\sum_{a,c}|\rho_{ac}|[\alpha_{ca}^{(01)''}+\alpha_{ca}^{(10)''}]\cos\phi_{ac}(T).
\end{align}
Finally, the total pulse energy change (narrowband and broadband components) can be obtained from Eqs. (\ref{eq:SlN3})  
\begin{align}\label{eq:SlN4}
S_{LH}^{(E)}(\omega_1,T)=-\frac{2}{\hbar^2}\sum_{a,c}|\rho_{ac}|\left([\beta_{ca}^{(01)'}+\beta_{ca}^{(10)'}]\sin\phi_{ac}(T)-[\beta_{ca}^{(01)''}+\beta_{ac}^{(10)''}]\cos\phi_{ac}(T)\right).
\end{align}

\subsection{Quadratic Probe}

The diagrams for the quadratic signal are depicted in Fig.~\ref{fig:diag}(d). We read the signal from the diagrams in Hilbert space (corresponding Liouville space expressions can be easily derived by expanding the loop diagram in the set of ladder diagrams) and obtain
\begin{align}\label{eq:Sqi31}
&S_{Q}^{(fd)}(\omega,t_0,\tau_0)=2\hbar\mathcal{I}\int_{-\infty}^{\infty}dt\int_{-\infty}^tdt'\int_{-\infty}^{\infty}dt_1\int_{-\infty}^{t_1}dt_1'\tilde{\mathcal{E}}^{*}(\omega)\mathcal{E}(t'-t_0)e^{i(\omega+\omega_0)(t-t_0)}\notag\\
\times[&\mathcal{E}^{*}(t_1-t_0)\mathcal{E}(t_1'-t_0)\langle \psi_i|G^{\dagger}(t-\tau_0)VG(t-t')V^{\dagger}G(t'-t_1)VG(t_1-t_1')V^{\dagger}G(t_1'-\tau_0)|\psi_i\rangle\notag\\
+&\langle \mathcal{E}(t_1-t_0)\mathcal{E}^{*}(t_1'-t_0)\psi_i|G^{\dagger}(t_1'-\tau_0)VG^{\dagger}(t_1-t_1')V^{\dagger}G^{\dagger}(t-t_1)VG(t-t')V^{\dagger}G(t'-\tau_0)|\psi_i\rangle],
\end{align}
where the first (second) term in Eq.~(\ref{eq:Sqi31}) represents the left (right) diagram in Fig.~\ref{fig:diag}(d).  The corresponding photon number and energy change signals are given by Eqs.~(\ref{eq:SN0}) and (\ref{eq:SE0}) respectively.

\subsubsection{Quadratic Broadband (QB probe)}

For a broadband pulse we expand the signal (\ref{eq:Sqi31}) in eigenstates which yields a compact formula
\begin{align}\label{eq:Sqifd710}
S_{QB}^{(fd)}(\omega,T)=-\mathcal{I}\frac{2i}{\hbar^4}\sum_{a,c,d}[\alpha_{cd}^{(\mathcal{E}\mathcal{E})}(\omega)\alpha_{da}^{(\mathcal{E}\mathcal{E})}-\alpha_{da}^{(\mathcal{E}\mathcal{E})}(\omega)\alpha_{dc}^{(\mathcal{E}\mathcal{E})*}]\rho_{ac}(T),
\end{align}
where $\rho_{ac}(T)=|\rho_{ac}|e^{-i\phi_{ac}(T)}$. In the off-resonant case $\alpha''=0$ and Eq.~(\ref{eq:Sqifd710}) reduces to Eq.~(\ref{eq:Sqi1}).
The integrated photon number signal (\ref{eq:SN0}) reads
\begin{align}
S_{QB}^{(N)}(T)=-\frac{4}{\hbar^4}\sum_{a,c,d}|\rho_{ac}|\alpha_{cd}^{(\mathcal{E}\mathcal{E})''}[\alpha_{da}^{(\mathcal{E}\mathcal{E})'}\sin\phi_{ac}(T)-\alpha_{da}^{(\mathcal{E}\mathcal{E})''}\cos\phi_{ac}(T)],
\end{align}
which clearly vanishes in the off-resonant case when $\alpha''=0$. The total energy change of the pulse (\ref{eq:SE0}) is given by
\begin{align}\label{eq:SqiE710}
S_{QB}^{(E)}(T)=-\mathcal{I}\frac{2i}{\hbar^4}\sum_{a,c,d}[\beta_{cd}^{(\mathcal{E}\mathcal{E})}\alpha_{da}^{(\mathcal{E}\mathcal{E})}-\beta_{da}^{(\mathcal{E}\mathcal{E})}\alpha_{dc}^{(\mathcal{E}\mathcal{E})*}]\rho_{ac}(T).
\end{align}

 \subsubsection{Quadratic Hybid (QH) probe}

The frequency-dispersed transmission of the broadband component of a hybrid broad-narrow probe (Eq.~(\ref{eq:Sqi31})) reads
\begin{align}\label{eq:Sqifd71H10}
S_{QH}^{(fd)}(\omega-\omega_1,\omega_1,T)=-\mathcal{I}\frac{2i}{\hbar^4}\sum_{a,c,d}[\alpha_{cd}^{(01)}(\omega)\alpha_{da}^{(10)}-\alpha_{da}^{(01)}(\omega)\alpha_{dc}^{(01)*}]\rho_{ac}(T),
\end{align}
which is similar to Eqs. (9)-(10) of \cite{zha14} for the model with zero linewidth of electronic states. The integrated photon number signal (\ref{eq:SN0}) which includes the change in both broadband and narrowband fields is
\begin{align}\label{eq:SqiN71H0}
&S_{QH}^{(N)}(\omega_1,T)=-\mathcal{I}\frac{2i}{\hbar^4}\sum_{a,c,d}[\alpha_{cd}^{(01)}\alpha_{da}^{(10)}+\alpha_{cd}^{(10)}\alpha_{da}^{(01)}-\alpha_{da}^{(01)}\alpha_{dc}^{(01)*}-\alpha_{da}^{(10)}\alpha_{dc}^{(10)*}]\rho_{ac}(T)
\end{align}
and the energy change of the pulse (Eq.~(\ref{eq:SE0})) is given by
\begin{align}\label{eq:SqiE71H0}
S_{QH}^{(E)}(\omega_1,T)=-\mathcal{I}\frac{2i}{\hbar^4}\sum_{a,c,d}[\beta_{cd}^{(01)}\alpha_{da}^{(10)}+\beta_{cd}^{(10)}\alpha_{da}^{(01)}-\beta_{da}^{(01)}\alpha_{dc}^{(01)*}-\beta_{da}^{(10)}\alpha_{dc}^{(10)*}]\rho_{ac}(T).
\end{align}

\subsection{Discussion of Resonant Signals}
\begin{figure}
\centering
\includegraphics[width=.7\textwidth]{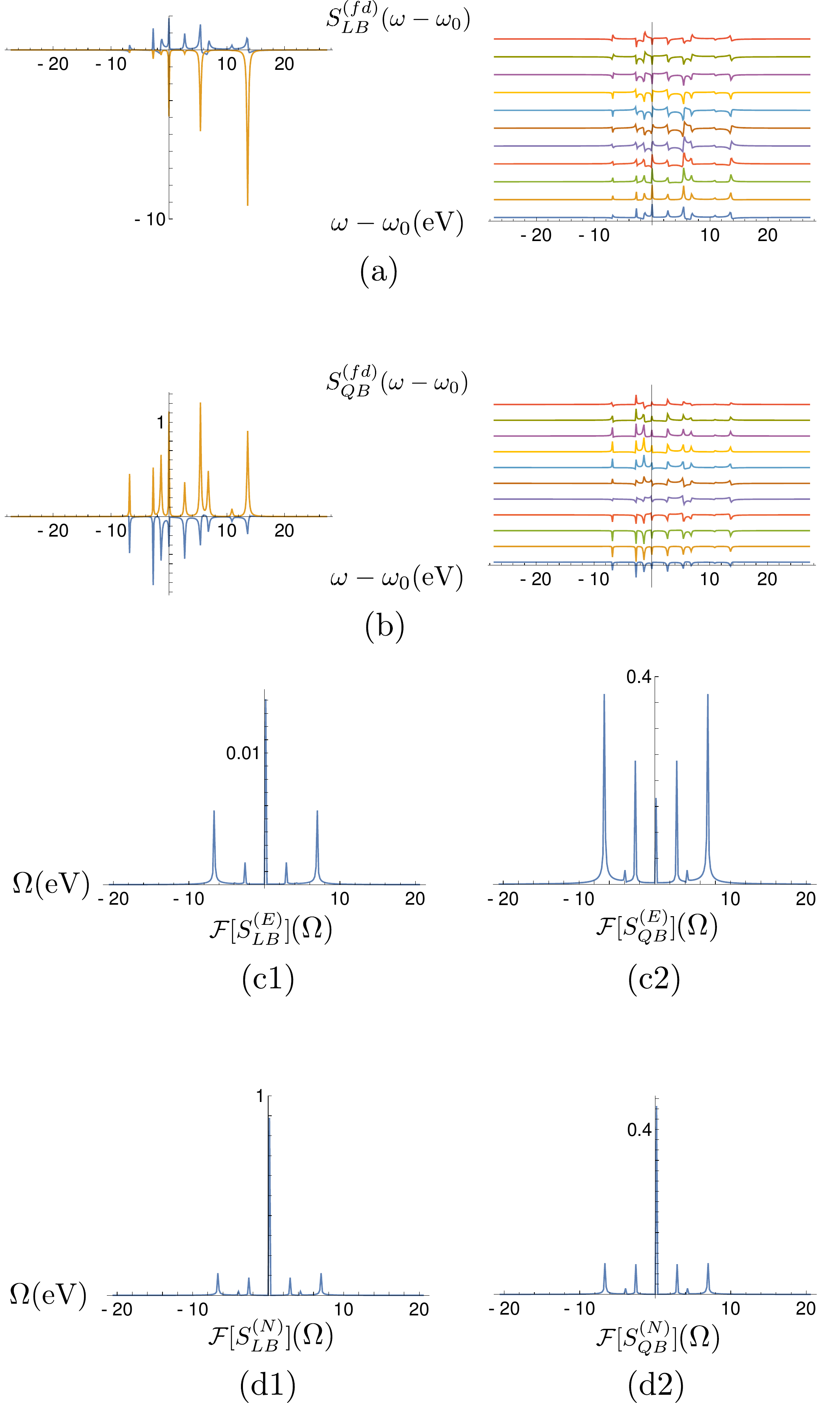}
\caption{(Color online) (a): Resonant linear broadband frequency-dispersed signal for the populations (time-independent) and coherences at $T=0$ (left) and time evolution of coherences (right). (b): Resonant quadratic broadband frequency-dispersed signal for the populations (time-independent) and coherences at $T=0$ (left) and time evolution of coherences (right). For coherences in both (a) and (b), $T$ advances in units of $\sim$60 attoseconds as one goes up the vertical axis.  (c1) and (c2): Resonant linear (c1) and quadratic (c2) broadband energy signal Fourier transformed.  (d1) and (d2): Resonant linear (d1) and quadratic (d2) broadband photon number signal Fourier transformed.}
\label{fig:ResLQB}
\end{figure}

For the simulations of the resonant signals, we extended the model used in the off-resonant case to include core states with energies $\varepsilon_{x(y)}\in\left\{ {136,141.5,149.5}\right\}(\text{eV})$. Figure \ref{fig:ResLQB} shows the resonant linear and quadratic broadband signals.  The frequency-dispersed signals (a/b) contain the valence-core resonances but not the Raman (this is due to the lack of field resolution).  However, the Raman transitions $\omega_{ac}$ are in the same region as the shifted valence-core transitions $\omega_{xa}-\omega_0$ so they appear similar. Since the $\omega_{xa}$ peaks can arise from any $a$, $c$ pair in the summation, the phases $\phi_{ac}^{\rho}$ are not directly accessible as the phases of peak oscillations.  Taking Fourier transform of the energy of photon number signals (c) and (d) gives peaks corresponding to the Raman transitions only since these are the oscillation frequencies.  As in the off-resonant case, the linear and quadratic signals have different relative peak intensities controlled by the different forms of the coupling.    
\par
\begin{figure}
\centering
\includegraphics[width=.8\textwidth]{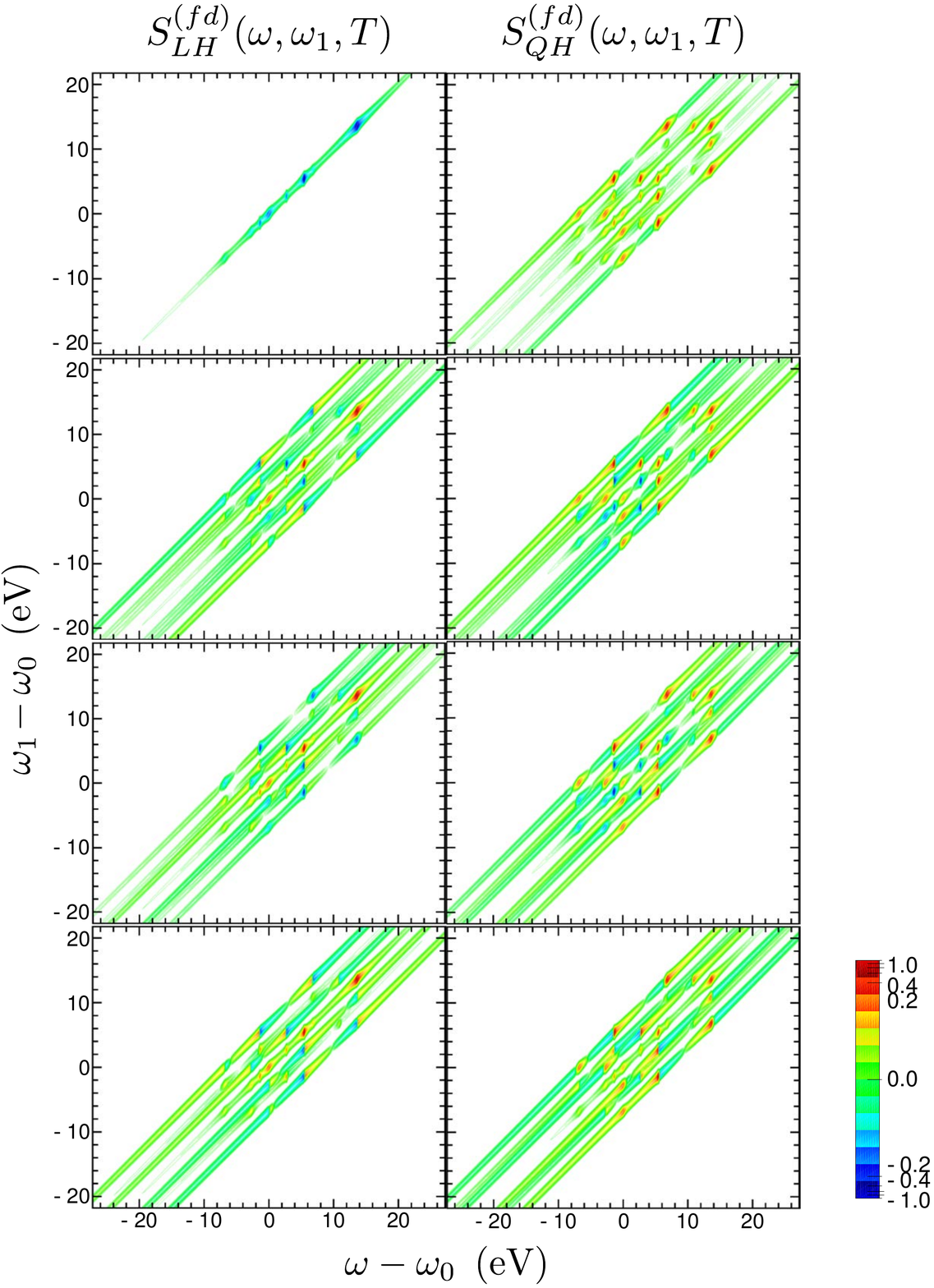}
\caption{(Color online) Resonant linear (left column) and quadratic (right column) hybrid frequency-dispersed signals.  The top of each column is the time-independent contribution due to populations.  The second panel from the top is the total signal at $T=0$ and the remaining two panels per column are for $T=240$(as) and $T=480$(as). This form of the signal allows one to disentangle valence-core ($\omega_{xa}$) transitions from Raman ($\omega_{ca}$) transitions.}
\label{fig:ResLQHfd}
\end{figure}

Figure \ref{fig:ResLQHfd} shows 2D spectra of the resonant linear (left column) and quadratic (right column) hybrid frequency-dispersed signals for populations only (top panels) and populations and coherences together at different times (remaining panels).  As $\omega_1-\omega_0$ varies, peaks corresponding to Raman transitions ($\omega_{ca}$) move along diagonals forming resonant streaks.  In contrast, the core transitions $\omega_{xa}$ do not vary with $\omega_{1}$ and therefore, each transition forms a series of repeated vertical peaks where the $\omega_{xa}$ transition intersects the diagonal streaks.  The magnitude of these peaks then reveals the strength of the coupling between the core and Raman transitions.  Note that the population contribution makes only a single diagonal streak in the linear case but results in all streaks in the quadratic case (this is due to the summation over the intermediate state $d$).  For the linear signal, we can therefore identify each linear streak with a particular $a$, $c$ pair in the summation and each streak will undergo Stokes/anti-Stokes oscillations at the respective phase $\Phi_{ac}(T)$, therefore allowing the retrieval of the phases $\phi_{ac}^{\rho}$.
\par
\begin{figure}
\centering
\includegraphics[width=.75\textwidth]{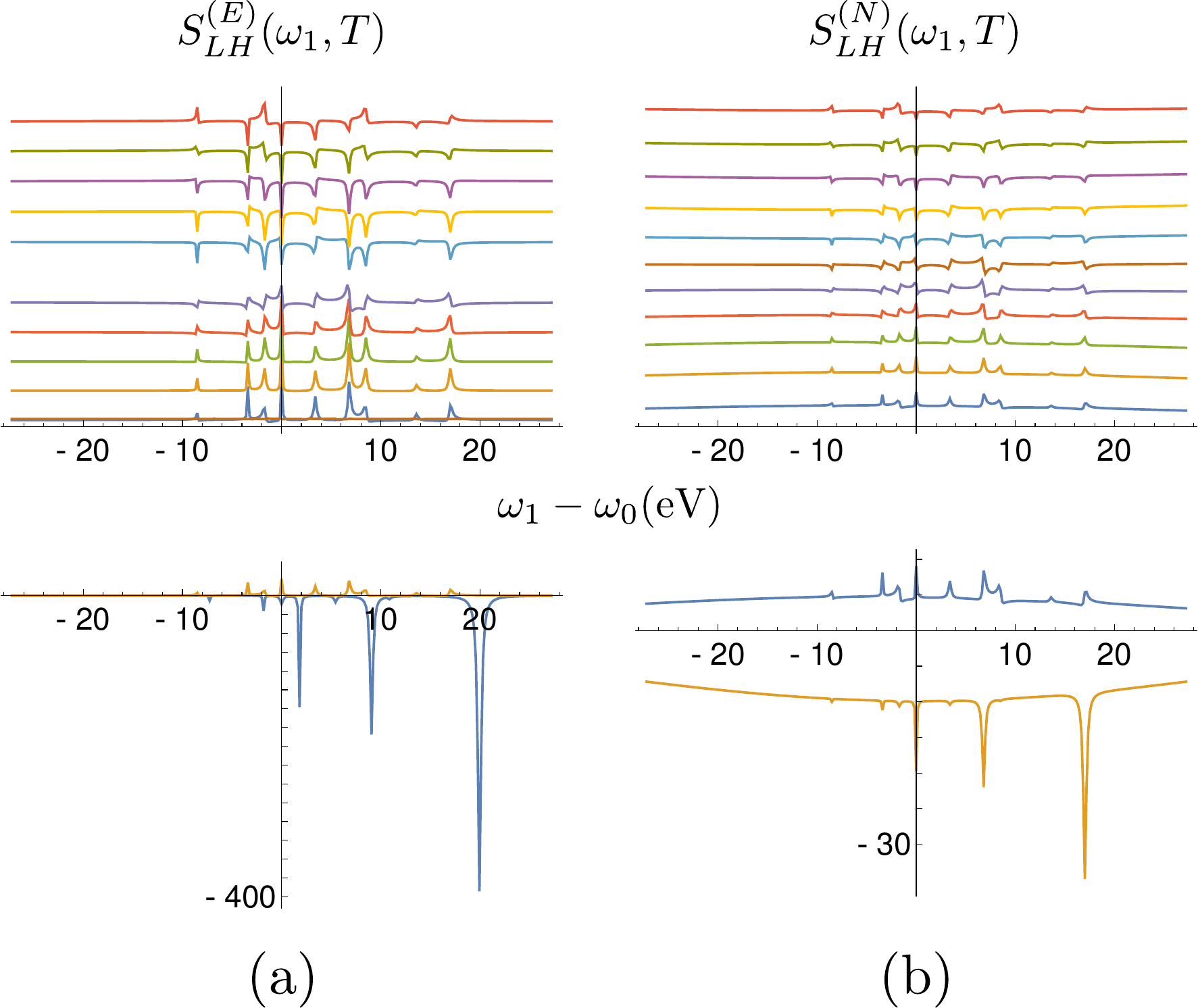}
\caption{(Color online) (a): Resonant linear hybrid energy signal for coherences at multiple times $T$ (top) and comparison of populations with initial coherences (bottom). (b): Resonant linear hybrid photon number signal for coherences at multiple times $T$ (top) and comparison of populations with initial coherences (bottom). For time-dependence of coherences, $T$ advances in units of $\sim$60 attoseconds as one goes up the vertical axis.}
\label{fig:ResLQHEN}
\end{figure}
\begin{figure}
\centering
\includegraphics[width=\textwidth]{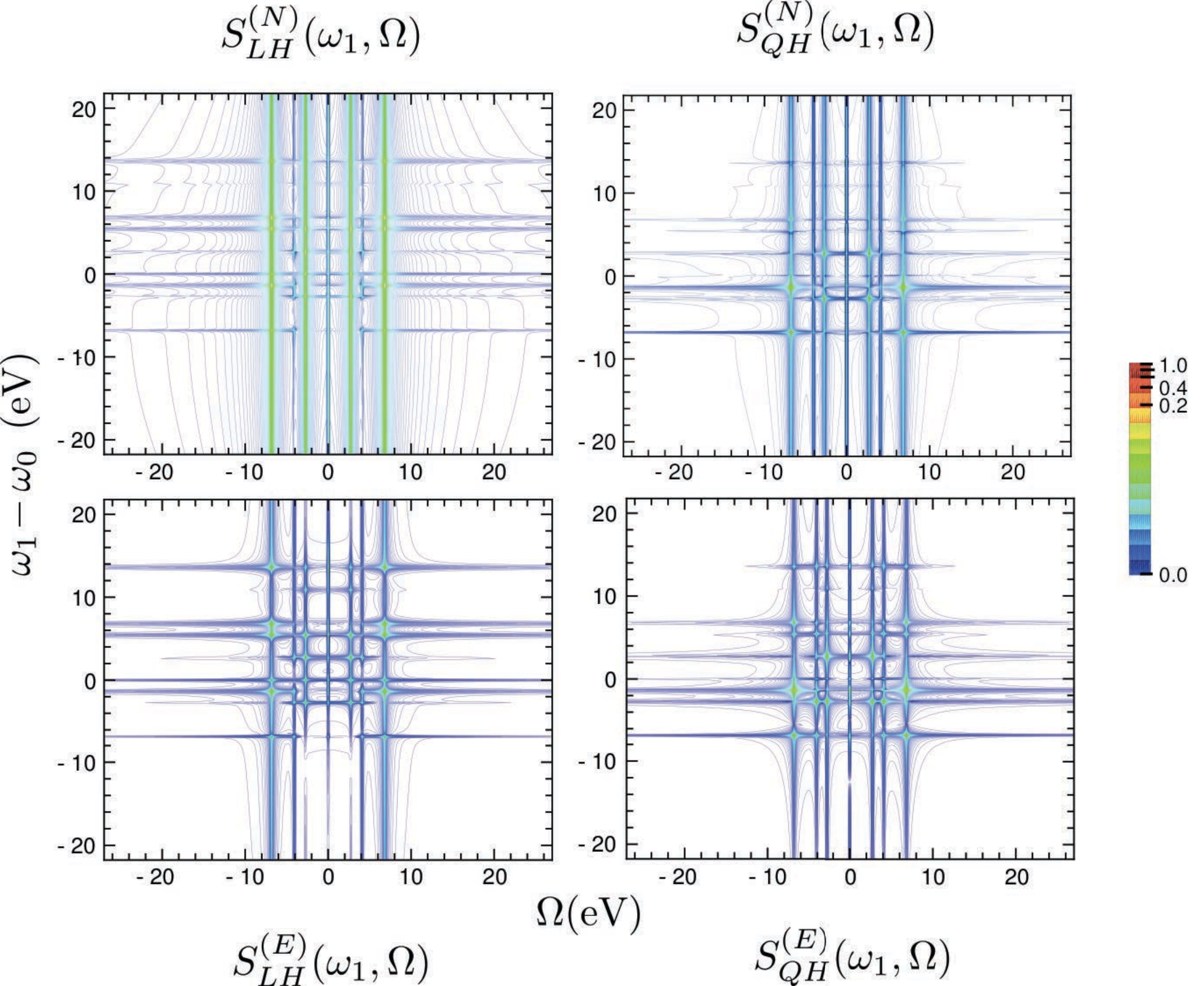}
\caption{(Color online) (Resonant linear (left column) and quadratic (right column) hybrid photon number (top) and energy change (bottom) signals.  These signals show $\Omega$ resonances at the Raman transitions $\omega_{ac}$ allowing their separation from the valence-core transitions $\omega_{xa}$.}
\label{fig:ResLQHEN2d}
\end{figure}

Figure \ref{fig:ResLQHEN} displays linear and quadratic hybrid energy and photon number signals.  In both cases, the contribution due to populations is stronger than that from coherences.  This is due to the resonant nature of the signal since the population contribution vanishes for linear off-resonant signals. Unlike in the off-resonant case, the signal contains sharp peaks due to the valence-core transitions but the same lack of field resolution prevents direct identification of the Raman transitions.  These can however be obtained via the Fourier transform.  Figure \ref{fig:ResLQHEN2d} shows the Fourier transform (magnitude) with respect to delay time $T$ of the linear and quadratic photon number and energy signals.  The quadratic is notably weaker but all four signals show the same basic pattern of $\omega_{xa}-\omega_0$ resonances along the $y$-axis and $\Omega=\omega_{ac}$ resonances along the $x$-axis. In Appendix \ref{app:ResAct}, we consider the resonant signals in the context of a nonstationary state prepared by an actinic pulse.

\section{Conclusions}
We presented a systematic classification scheme applicable to spectroscopic techniques in which a nonstationary state's creation is well-seperated from it's detection (pump-probe style spectroscopies).  In this scheme, a particular spectroscopic technique is specified by the choice of state preparation and detection procedures. We consider various detection procedures and examine what sort of information one may obtain and in what ways the signals considered differ. We demonstrated the various signals for a model consisting of a few valence and a few core states and calculated the frequency-dispersed, photon number, and energy change signals for both linear or quadratic field intensity scaling and broad or hybrid spectral field shapes.  

There are two mechanisms whereby the photon modes can change their occupation numbers.  A photon can either redistribute amongst the modes, swapping from one frequency to another as in Raman interactions or, alternatively, it can be absorbed or emitted by matter.  The former process is governed by the real parts of the polarizabilities $\alpha$ (and $\beta$) while the latter is governed by the imaginary parts.  In the off-resonant regime, only the redistributive mechanism is operative.  In the resonant regime, both contribute and oscillate with different phases in delay time $T$ (the redistributive being a Sine funcion and the absorbtive/emittive being a Cosine).  
\par
The spectrum of the nonstationary state (the energy levels of the occupied states), can be obtained by taking the Fourier transform of any of the time-dependent signals, resonant or off-resonant.  The relative magnitude of the various peaks in these signals varies depending on the coupling constants for each signal.  In the simpler, linear signals, the relative peak magnitudes are directly indicative of the product of the coherence $\rho_{ac}$ and the polarizability $\alpha_{ac}$, while in the quadratic case, there is an additional $\alpha$ and a free summation over the intermediate state. Using broadband pulses, we had to sacrifice the time resolution to obtain frequency resolution of the $\omega_{ac}$ transitions.  This limitation was surpassed by using a shaped, hybrid pulse consisting of broad and narrowband components.  The resultant frequency-dispersed signals naturally show the peaks at the $\omega_{ac}$ transitions which oscillate in time.  In the linear case, each peak oscillates with its own phase $\omega_{ac}T+\phi_{ac}^\rho$, thus allowing determination of the phase induced by the state preparation process.  This technique therefore exploits the simultaneous resolution of narrowband and detected frequencies combined with the time-resolution granted by the broadband pulse to obtain the phases of the electronic wavepacket.  Without frequency-dispersed detection, one is left with the photon number and energy change signals.  Even though they contain less information than the frequency dispersed hybrid signals, the resolution granted by the narrowband pulse allows one to access essentially the same information as the frequency dispersed broadband signal.  
\par
Finally, we note that the present formalism also applies to optical signals. X-ray signals allow one to probe valence excitations rather than vibrations. Furthermore, both the initial preparation and/or the preparation process can then be the product of an X-ray scattering, photoionization, or Auger process (as recently discussed in \cite{bennett2015probing,mcfarland2014ultrafast}) in addition to an X-ray Raman process. The signals obtained by a Raman probe then detect both the amplitude and phase of the coherent superpositions of singly or doubly ionized states. These are signatures of many- body effects in the photoionization and the Auger processes.
\par
Compared to our previous work, here we present the most general systematic description of series of X-ray Raman signals. Using abbreviations of the signals of the present paper in our earlier work \cite{dor132}  we presented $S_{QH}^{(fd)}$, $S_{Q12}^{(N)}$, and partially $S_{QH}^{(N)}$ signals. Using the present terminology, the signals  studied in \cite{aga15} are $S_{QH}^{(fd)}$ and $S_{LH}^{(fd)}$. Finally, in the most recent work \cite{batignani2015energy}, we investigated $S_{QH}^{(N)}$ signal.

\appendix

\section{Multidimensional Off-Resonant Signals Initiated by an Impulsive Raman Process}\label{app:MultiRaman}

\subsection{Two-pulse Time-Domain Experiment}

We now turn to a more elaborate experiment that probes the same material quantity described by Eqs.~(\ref{eq:Sqi1}) but involves two broadband pulses separated in time by delay $T_2$. At time $\tau_0$  the system is prepared in $\rho_{ac}$. Following the delay $T_1=t_1-\tau_0$ the first probe pulse $\mathcal{E}_1$ centered at $t_1$  interacts with the system and after another delay $T_2=t_2-t_1$ the second pulse $\mathcal{E}_2$ centered at $t_2$ yields the two-dimensional spectra as a function of the two time delays. The corresponding diagrams are given in Fig. \ref{fig:diag}(b). The difference is in the detection as well as the fact that the signal is linear in intensity of $\mathcal{E}_1$ as well as $\mathcal{E}_2$. The frequency dispersed transmission is a 3D signal 
\begin{align}
&S_{Q12}^{(fd)}(\omega,T_1,T_2)=\frac{2}{\hbar^2}\int\frac{d\omega_1}{2\pi}|\mathcal{E}_2(\omega)||\mathcal{E}_1(\omega)|\sum_{a,c,d}|\rho_{ac}|\alpha_{cd}^{(0)}\alpha_{da}^{(0)}\notag\\
&\times[|\mathcal{E}_2(\omega-\omega_{ad})||\mathcal{E}_1(\omega_1-\omega_{dc})|\cos(\omega_{ad}T_2+\phi_{ac}(T_1))\notag\\
&-|\mathcal{E}_2(\omega-\omega_{dc})||\mathcal{E}_1(\omega_1-\omega_{ad})|\cos(\omega_{dc}T_2+\phi_{ac}(T_1))],
\end{align}
the number photon signal $S_{Q12}^{(N)}(T_1,T_2)=0$ vanishes, and the energy change signal reads
\begin{align}\label{eq:SQ12E}
&S_{Q12}^{(E)}(T_1,T_2)=-\frac{2}{\hbar^2}\int\frac{d\omega}{2\pi}\int\frac{d\omega_1}{2\pi}|\mathcal{E}_2(\omega)||\mathcal{E}_1(\omega)|\sum_{a,c,d}|\rho_{ac}|\alpha_{cd}^{(0)}\alpha_{da}^{(0)}\notag\\
&\times[\hbar\omega_{da}|\mathcal{E}_2(\omega-\omega_{ad})||\mathcal{E}_1(\omega_1-\omega_{dc})|\cos(\omega_{ad}T_2+\phi_{ac}(T_1))\notag\\
&+\hbar\omega_{dc}|\mathcal{E}_2(\omega-\omega_{dc})||\mathcal{E}_1(\omega_1-\omega_{ad})|\cos(\omega_{dc}T_2+\phi_{ac}(T_1))].
\end{align}
This is a sum of two fluxes $a\to d$ and $c\to d$. Note, that compared to the signal (\ref{eq:SE5}), the two-pulse signal (\ref{eq:SQ12E})  carries information about the phase and dynamics of the system after interacting with the probe $\mathcal{E}_1$ during the delay $T_2$. This is not accessible by the QB signal.

\subsection{Multidimensional Stimulated Raman Spectroscopy with an Off-Resonant Actinic Excitation}

There are multiple ways to prepare initial state $\rho_{ac}$ superposition. These include photoionization, Auger process, off-resonant or resonant impulsive Raman excitation. So far we did not specify the preparation. In the standard formulation of multidimensional spectroscopy,  the molecule is prepared by a series of short temporally well-separated  pulses $\mathcal{E}_j$ at time $\tau_j$, $j=1,2,...$. 
Followed by linear transmission of the probe (assuming no dynamics between preparation time $\tau_0$ and interaction with probe at $t$) one can then expand the correlation function in Eqs. (\ref{eq:Str0}) - (\ref{eq:SE0})
\begin{align}
\langle\langle I| \alpha^{(0)}_L(t)&e^{-\frac{i}{\hbar}\int H'_-(\tau)d\tau}|\rho_i(\tau_0)\rangle\rangle=|\mathcal{E}_0(\tau_0)|^2\langle [\alpha(t),\alpha(\tau_0)],\rho_g\rangle\notag\\
&-\frac{i}{\hbar}|\mathcal{E}_0(\tau_0)|^2|\mathcal{E}(\tau_1)|^2[[\alpha(t),\alpha(\tau_0)|,\alpha(\tau_1)],\rho_g\rangle\notag\\
&+\left(-\frac{i}{\hbar}\right)^2|\mathcal{E}_0(\tau_0)|^2|\mathcal{E}(\tau_1)|^2|\mathcal{E}(\tau_2)|^2[[[\alpha(t),\alpha(\tau_0)|,\alpha(\tau_1)],\alpha(\tau_2)],\rho_g\rangle+...
\end{align}
The same result can be obtained by neglecting the preparation (at $\tau_0$ system is in  $\rho_g$) and expanding the dynamics between preparation time $\tau_0$ and interaction with the probe pulse at $t$ assuming series of short pulses prior to the probe pulse.

In many X-ray spectroscopy applications, a single resonant actinic pulse $\mathcal{E}_A$ prepares the system in the superposition of valence states. We consider an off-resonant preparation. We do not expand the signal in sum over states and obtain a general expression for the initial state described by the density matrix $\rho_i(\tau_0)$ assuming that the preparation process is temporally well separated from the probe 
\begin{align}
|\rho_i(\tau_0)\rangle\rangle=\frac{-i}{\hbar}\int_{-\infty}^{\infty}dt'' |\mathcal{E}_A(t''-\tau_0)|^2\alpha^{(0)}_-(t'')|\rho_g\rangle\rangle.
\end{align}
The linear frequency-dispersed transmission signals (\ref{eq:Str11}) then reads
\begin{align}\label{eq:Str113}
S_L^{(fd)}(\omega,t_0,\tau_0)=\mathcal{I}\frac{-2i}{\hbar^2}\int_{-\infty}^{\infty}dt\int_{-\infty}^{\infty}dt'e^{i\omega(t-t_0)}\mathcal{E}^{*}(\omega)\mathcal{E}(t-t_0)|\mathcal{E}_A(t'-\tau_0)|^2\langle[\alpha^{(0)}(t),\alpha^{(0)}(t')],\rho_g\rangle,
\end{align}
\begin{align}\label{eq:SN113}
S_L^{(N)}(\omega,t_0,\tau_0)=\mathcal{I}\frac{-2i}{\hbar^2}\int_{-\infty}^{\infty}dt\int_{-\infty}^{\infty}dt'|\mathcal{E}(t-t_0)|^2|\mathcal{E}_A(t'-\tau_0)|^2\langle[\alpha^{(0)}(t),\alpha^{(0)}(t')],\rho_g\rangle,
\end{align}
\begin{align}\label{eq:SE113}
S_L^{(E)}(\omega,t_0,\tau_0)=-\mathcal{I}\frac{2}{\hbar}\int_{-\infty}^{\infty}dt\int_{-\infty}^{\infty}dt'\dot{\mathcal{E}}^{*}(t-t_0)\mathcal{E}(t-t_0)|\mathcal{E}_A(t'-\tau_0)|^2\langle[\alpha^{(0)}(t),\alpha^{(0)}(t')],\rho_g\rangle,
\end{align}
The quadratic frequency-dispersed transmission signal (\ref{eq:Str411})  is
\begin{align}\label{eq:Sqtr113}
S_Q^{(fd)}(\omega,t_0,\tau_0)&=\mathcal{I}\frac{-2}{\hbar^3}\int_{-\infty}^{\infty}dt\int_{-\infty}^{t}dt'\int_{-\infty}^{\infty}dt''e^{i\omega(t-t_0)}\mathcal{E}^{*}(\omega)\mathcal{E}(t-t_0)|\mathcal{E}(t'-t_0)|^2|\mathcal{E}_A(t''-\tau_0)|^2\notag\\
&\times\langle[[\alpha^{(0)}(t),\alpha^{(0)}(t')],\alpha^{(0)}(t'')]\rho_g\rangle.
\end{align}
\begin{align}\label{eq:SqN113}
S_Q^{(N)}(t_0,\tau_0)&=\mathcal{I}\frac{-2}{\hbar^3}\int_{-\infty}^{\infty}dt\int_{-\infty}^{t}dt'\int_{-\infty}^{\infty}dt''|\mathcal{E}(t-t_0)|^2|\mathcal{E}(t'-t_0)|^2|\mathcal{E}_A(t''-\tau_0)|^2\notag\\
&\times\langle[[\alpha^{(0)}(t),\alpha^{(0)}(t')],\alpha^{(0)}(t'')]\rho_g\rangle.
\end{align}
\begin{align}\label{eq:SqE113}
S_Q^{(E)}(t_0,\tau_0)&=\mathcal{I}\frac{2i}{\hbar^2}\int_{-\infty}^{\infty}dt\int_{-\infty}^{t}dt'\int_{-\infty}^{\infty}dt''\dot{\mathcal{E}}^{*}(t-t_0)\mathcal{E}(t-t_0)|\mathcal{E}(t'-t_0)|^2|\mathcal{E}_A(t''-\tau_0)|^2\notag\\
&\times\langle[[\alpha^{(0)}(t),\alpha^{(0)}(t')],\alpha^{(0)}(t'')]\rho_g\rangle.
\end{align}
These agree with original expressions obtained for homodyne detection \cite{tan93}. The photon number and energy change signals can be obtained similarly.  They provide a different gating of the same response function given by multiple commutators of the bare polarizability $\alpha^{(0)}$.

\section{Off-Resonant Limit of the Resonant Polarizability}

We now examine the frequency-dispersed signal (\ref{eq:Slfdbb1}) and show that it reduces to the off-resonant result of Eq. (\ref{eq:Str2}). We first note that in the off-resonant case (Eqs. (\ref{eq:def1})) we can define the following quantity:
\begin{align}\label{eq:casub}
\alpha_{ca}^{(\mathcal{E}\mathcal{E})}(\omega)=\mathcal{E}^{*}(\omega)\mathcal{E}(\omega+\omega_{ac})\alpha_{ca}^{(0)},
\end{align}
where
\begin{align}\label{eq:aoffres}
\alpha_{ca}^{(0)}=\sum_x\frac{(\mathbf{e}_\mathcal{E}\cdot V_{ax})(\mathbf{e}_\mathcal{E}\cdot V_{xc})}{-\omega_{xa}}
\end{align}
is the off-resonant polarizability. Choosing real basis set wave functions, the dipole moments are real. In this case the off-resonant polarizability (\ref{eq:aoffres}) is real as well.  Note, that technically this polarizability still contains field dependence via polarization of the probe pulse. Similarly
\begin{align}\label{eq:acsub}
\alpha_{ca}^{(\mathcal{E}\mathcal{E})}(\omega)=\mathcal{E}^{*}(\omega)\mathcal{E}(\omega-\omega_{ac})\alpha_{ca}^{(0)}.
\end{align}
Substituting Eqs. (\ref{eq:casub}) - (\ref{eq:acsub}) into Eq. (\ref{eq:Slfdbb1}) we obtain (\ref{eq:Str2}). The other detection protocols and pulse shapes can be calculated similarly.

\section{Multidimensional  stimulated Raman spectroscopy with resonant actinic excitation}\label{app:ResAct}

\begin{figure}[h]
\centering
\includegraphics[width=.7\textwidth]{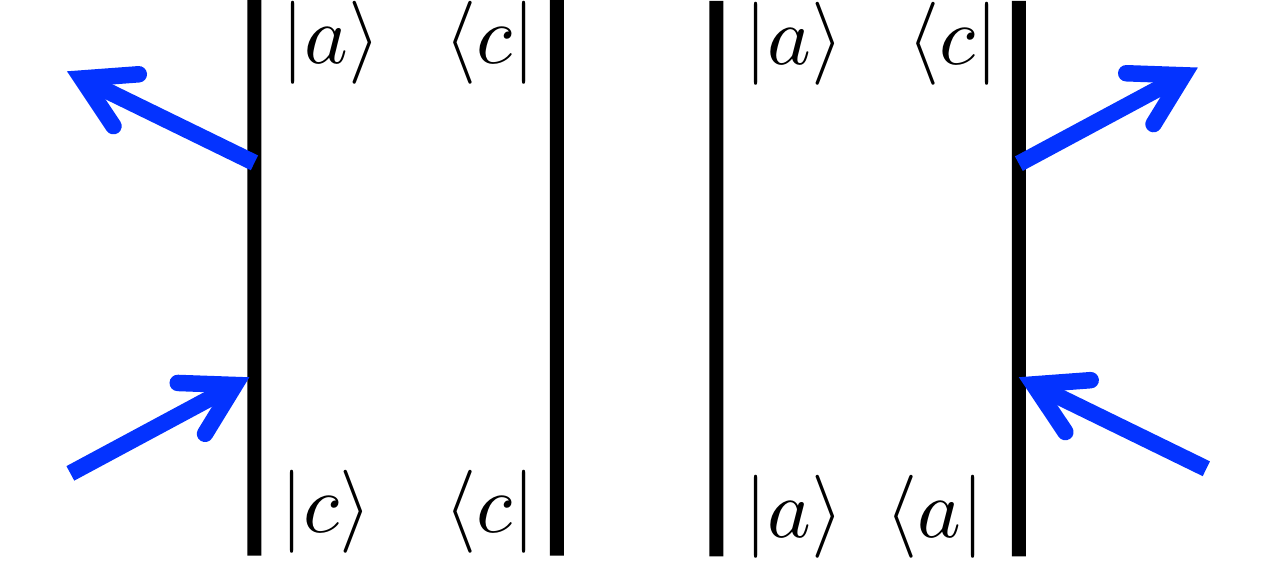}
\caption{(Color online) Set of ladder diagrams for resonant actinic preparation}
\label{fig:resprep}
\end{figure}

If the preparation involves resonant actinic pulse the  relevant excitation process is given by two diagrams shown in Fig. \ref{fig:resprep} and the straightforward calculation yields
\begin{align}\label{eq:rhoacRam}
\rho_{ac}=\frac{2}{\hbar^2}\alpha_{ac}^{(AA)''}.
\end{align}
We can now substitute this matrix element into the signals and obtain a compact form in terms of $\alpha$ only. For the case of broadband pulse Eqs. (\ref{eq:Slfdbb1}), (\ref{eq:SlN20}) and (\ref{eq:SEN20}) yield
\begin{align}\label{eq:Slfdbb11}
S_{LB}^{(fd)}(\omega,T)=-\frac{4}{\hbar^4}\sum_{a,c}\alpha_{ac}^{(AA)''}\left(\alpha_{ca}^{(\mathcal{E}\mathcal{E})'}(\omega)\sin\omega_{ac}T-\alpha_{ca}^{(\mathcal{E}\mathcal{E})''}(\omega)\cos\omega_{ac}T\right),
\end{align}
\begin{align}\label{eq:SlN201}
S_{LB}^{(N)}(T)=\frac{4}{\hbar^4}\sum_{a,c}\alpha_{ac}^{(AA)''}\alpha_{ca}^{(\mathcal{E}\mathcal{E})''}\cos\omega_{ac}T,
\end{align}
\begin{align}\label{eq:SEN201}
S_{LB}^{(E)}(T)=-\frac{2}{\hbar^4}\sum_{a,c}\alpha_{ac}^{(AA)''}\left[\hbar\omega_{ac}\alpha_{ca}^{(\mathcal{E}\mathcal{E})'}\sin\omega_{ac}T-(\beta_{ca}^{(\mathcal{E}\mathcal{E})''}+\beta_{ac}^{(\mathcal{E}\mathcal{E})''})\cos\omega_{ac}T\right],
\end{align}

\acknowledgments
The support of the Chemical Sciences, Geosciences, and Biosciences division, Office of Basic Energy Sciences, Office of Science, U.S. Department of Energy  as well as from the National Science Foundation (grant CHE-1361516), and the National Institutes of Health (Grant GM-59230) is gratefully acknowledged.  Kochise Bennett was supported by DOE.


%

\end{document}